\RequirePackage{silence}
\let\orilabel\label 
\documentclass[aps,pra,reprint,10pt,floatfix,superscriptaddress]{revtex4-2}
\let\label\orilabel

\usepackage[T1]{fontenc}
\usepackage[utf8]{inputenc}
\usepackage{cancel}
\usepackage{soul}
\usepackage{amsfonts}
\usepackage{overpic}
\usepackage[english]{babel}
\usepackage{graphicx}
\usepackage[dvipsnames]{xcolor}
\usepackage{hyperref}
\hypersetup{colorlinks,allcolors=blue}
\usepackage{amsmath}
\usepackage{babel}

\begin{document}
\title{Detuning-assisted cavity enhancement in spectroscopy}

\author{Subhasish Guha}
\email{subhasish.guha@warwick.ac.uk}
\affiliation{Department of Physics, University of Warwick, Coventry, CV4 7AL, United Kingdom}

\author{Sourav Das}
\affiliation{Department of Physics, University of Warwick, Coventry, CV4 7AL, United Kingdom}
\affiliation{D\'epartement de g\'enie physique, \'Ecole polytechnique de Montr\'eal, Montr\'eal, QC, H3T 1J4, Canada}

\author{Francesco Albarelli}
\affiliation{Università di Parma, Dipartimento di Scienze Matematiche, Fisiche e Informatiche, I-43124 Parma, Italy}
\affiliation{INFN—Sezione di Milano-Bicocca, Gruppo Collegato di Parma, I-43124 Parma, Italy}

\author{Aiman Khan}
\affiliation{Manufacturing Metrology Team, Faculty of Engineering, University of Nottingham, Nottingham NG7 2RD, United Kingdom}

\author{Elnaz Darsheshdar}
\affiliation{Department of Physics, University of Malta, Msida MSD 2080, Malta}

\author{Adam D. Wright}
\email{a.d.wright@warwick.ac.uk}

\author{Animesh Datta}
\affiliation{Department of Physics, University of Warwick, Coventry, CV4 7AL, United Kingdom}

\begin{abstract}
Cavity-enhanced spectroscopy is a widely used experimental technique. We show that detuning the cavity frequency from the transition of interest can enhance the precision for dipole moment estimation by up to two orders of magnitude beyond the improvement when the cavity is resonant with the transition. We show this for both classical continuous-wave (CW) and pulsed single-photon probes, even in the presence of losses. We trace this enhancement to a detuning-controlled redistribution of information between the light- and matter-like polaritons, with losses determining the optimal branch. We show that this enhancement remains accessible using conventional photodetection and time-correlated single-photon counting.
\end{abstract}
\maketitle

\section{Introduction}
Absorption spectroscopy is a widely used experimental technique for probing matter samples~\cite{Lackner07,Bakker02,Laj:09} by illuminating them with light of known intensity $I_0$ and detecting the output light. The output intensity measured at a particular frequency is dictated by Beer-Lambert's law as $I_{\mathrm{out}}=I_0\exp\left( -\alpha\left(\omega \right)L\right)$ where $\alpha\left(\omega \right)$ is the coefficient of absorption at frequency $\omega$ and $L$ is the path length of light through the sample. The smallest changes in the output light that can be reliably detected are limited by the accuracy of the detector and the source. Thus, schemes to lower the smallest detectable $\alpha\left(\omega \right)$ have often focussed on increasing the path length $L$~\cite{Herriott64, McManus95, White76}. This has been achieved using high finesse optical cavities~\cite{Kastler:74, OKeefe88} which increase the effective optical path length by trapping the light between the cavity mirrors. 
\begin{figure}[t]
    \centering
    \includegraphics[width=\linewidth]{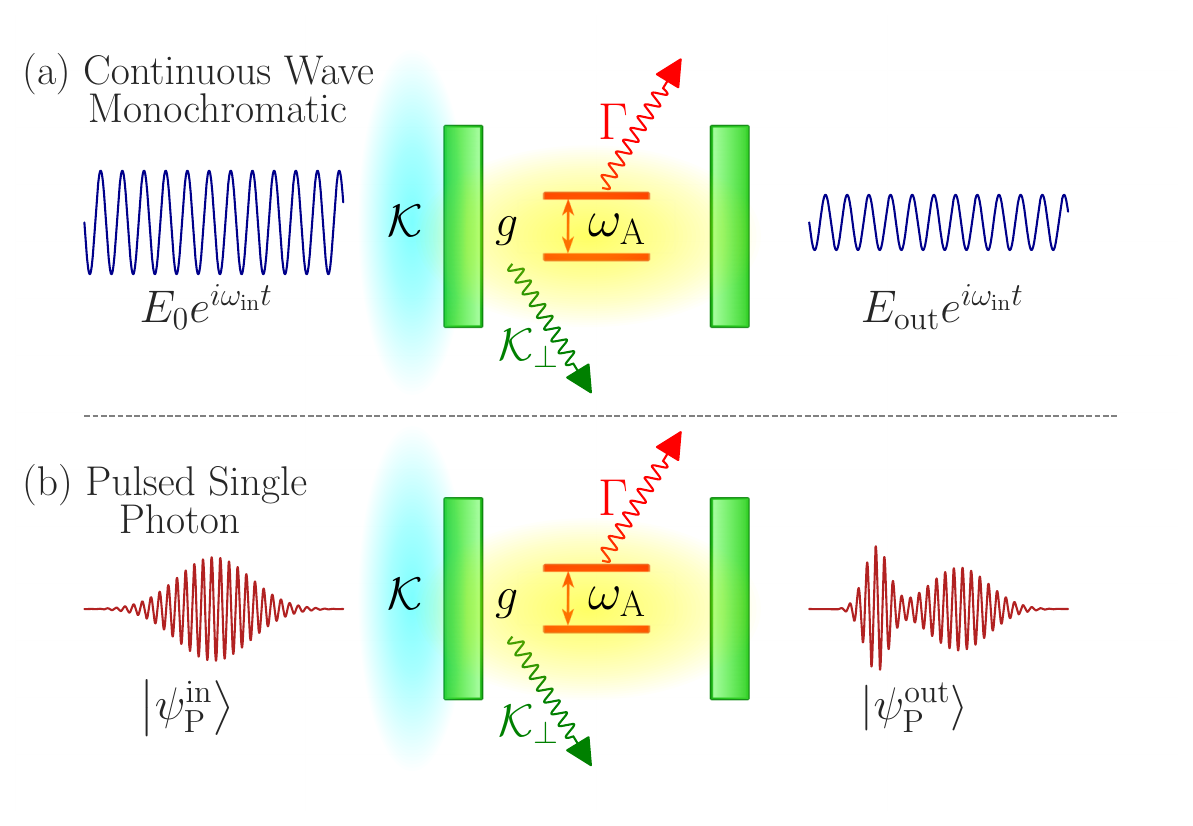}  
    
    \caption{Schematic of cavity-enhanced  spectroscopy, showing the two types of light probes we consider. In typical models for spectroscopy, continuous-wave monochromatic light probes (a) are considered, and the output light amplitude is modified according to the transmission of the sample.  For pulsed single photon probes (b), the sample modifies the shape of the input pulse. $\mathcal{K}$ represents the strength of coupling between the input probe light and the cavity mode, whereas $g$ is the interaction strength between the cavity and the two-level atom (TLA).  $\Gamma$ and $\mathcal{K}_{\perp}$ represent the losses from the atom and the cavity into the continuum respectively.    }
    \label{fig:Losses_schematic}
\end{figure}

Typical theoretical descriptions of cavity-enhanced spectroscopy (CES) are semiclassical~\cite{McKillop_Weichman_2025,Lehmann1996,gagliardi2014cavity,Fiedler2003}. There the probe light, considered to be an electromagnetic wave with amplitude $E_{0}$, is incident on a Fabry-Perot  cavity composed of two mirrors, which contains the sample. See Fig.~\ref{fig:Losses_schematic} (a). The incident light is taken to be continuous-wave (CW) and monochromatic at a particular probe frequency. While recent experiments have pushed CES towards quantum regimes~\cite{Stankiewicz:26,Ulanowski:25}, the widely adopted theoretical measure of cavity enhancement remains, via the semiclassical treatment, the effective path length $L_{\mathrm{eff}}\propto 2\omega_{\mathrm{C}}L/\mathcal{K}$, where $\omega_{\mathrm{C}}$ is the cavity mode frequency and $\mathcal{K}$ is the strength of coupling between the input probe light and the cavity mode.

In this work,  we show that for dipole moment estimation the physics of cavity QED can be leveraged to increase the cavity-enhancement by detuning the cavity mode from the transition frequency of the matter sample in consideration. We dub this detuning-assisted enhancement (DAE). This stands in contrast to other complicated schemes to increase the cavity-enhancement in CES~\cite{Maity:21,Gianfrani:24}. We show that for cavities operating in the strong coupling regime (where the atom-cavity interaction strength $g$ exceeds $\mathcal{K}$), DAE persists across a large range of values for the atomic loss  and could be a practical means of increasing the attainable precision for dipole moment estimation by over two orders of magnitude. The importance of the dipole moment lies in its role as the quantitative scaling factor required to translate measurements into physical insights  in various scenarios ranging from astrophysics to tests of fundamental physics~\cite{Terabayashi:26,Haq:10,Baig:07,Barker:68,Liang:24}. 

For conceptual clarity, we illustrate DAE for a sample which is a single two-level atom (TLA), where the dipole moment of the atomic transition is proportional to the atom-cavity coupling strength $g$. Most importantly, we show that DAE exists when the cavity is illuminated by a CW or single photon pulses (See Fig~\ref{fig:Losses_schematic}  (a), (b).) and the signatures of DAE are qualitatively similar in both cases. The former consists of a semi-classical drive, operating in the weak, unsaturated regime ensuring that it retains Poissonian statistics after the interaction. The latter considers a \emph{fully} quantum light-matter interaction with a pulsed single photon. While we consider DAE to be a quantum effect as it emerges from the polaritonic behaviour of the atom-cavity system, previous studies have argued that polaritonic phenonmena can be explained classically \cite{McKillop_Weichman_2025,MolecularPolaritons}. Nevertheless, our results show DAE may  be utitlised for both classical metrology and emerging quantum sensing applications. 

Our analysis is also valid for atomic ensembles when the cavity is probed with single-photon pulses or weak CW light, since the ensembles can be mapped to effective single atom models under the assumptions of the Tavis-Cummings model in the single excitation sector~\cite{Tavis:68}. 

We arrive at our conclusions by surrendering $L_{\mathrm{eff}}$ as the theoretical measure of cavity-enhancement. Rather, we  recast CES as a parameter estimation problem that seeks to estimate some sample (matter) parameter $\theta$. In estimation theory, the Fisher information (FI), which sets a lower bound on the variance of any unbiased estimator for $\theta$~\cite{Rao_1992,cramer2016,FisherFlow}, arises as a natural measure of cavity-enhancement. We will refer to the inverse of the variance as precision and higher values of the FI will indicate an increase to maximum achievable precision. We note here that the FI is a dimensional quantity, with the FI of estimating the parameter $\theta$ having dimensions of $\theta^{-2}$. We thus focus on the dimensionless FI defined as the FI multiplied by $\theta^2$. 
A more detailed discussion on classical and quantum estimation theory is in Appendix~\ref{Est_Theory}.

This article is structured as follows. In section~\ref{SemiClassical}, we first introduce the typical semiclassical model used to describe CES and the transmission function that we obtain for this system. We present the results demonstrating DAE in different physically relevant parameter regimes. Section~\ref{Quantum} provides the quantum description of CES for pulsed single photon illumination. Within this fully quantum description, we analytically calculate the quantum state of the output light and the associated quantum Fisher information (QFI)~\cite{Paris2009quantum} for  dipole moment estimation. This allows us to establish DAE in the fully quantum regime. In section~\ref{Discussion}, we conclude with a discussion of our main results.

\section{Results}

We begin with a brief description of the semiclassical model and its estimation theoretic analysis. Mathematically, the dimensionless FI of estimating the parameter $\mathcal{G}=g^2$ can be shown to be equivalent to the dimensionless FI of estimating the square of the dipole moment. See Appendix~\ref{Est_Theory}. Thus, the FI of estimating $\mathcal{G}$ reflects the maximum attainable precision for estimating the dipole moment and other related atomic parameters such as the linewidth. The choice of $\mathcal{G}$ as the parameter to estimate also allows comparison with previous works which have studied spectroscopy within the estimation theory framework~\cite{Francesco2023,Sourav1,Chinni:26}.

We follow this with a quantum description of CES along with the quantum limits on the precision of estimating the dipole moment. This is given by the QFI. The QFI of estimating a parameter from a given quantum state is the maximum FI that can be attained by any measurement on that given state. 
\begin{figure*}[t]
    \centering
    \begin{overpic}[scale=0.58]{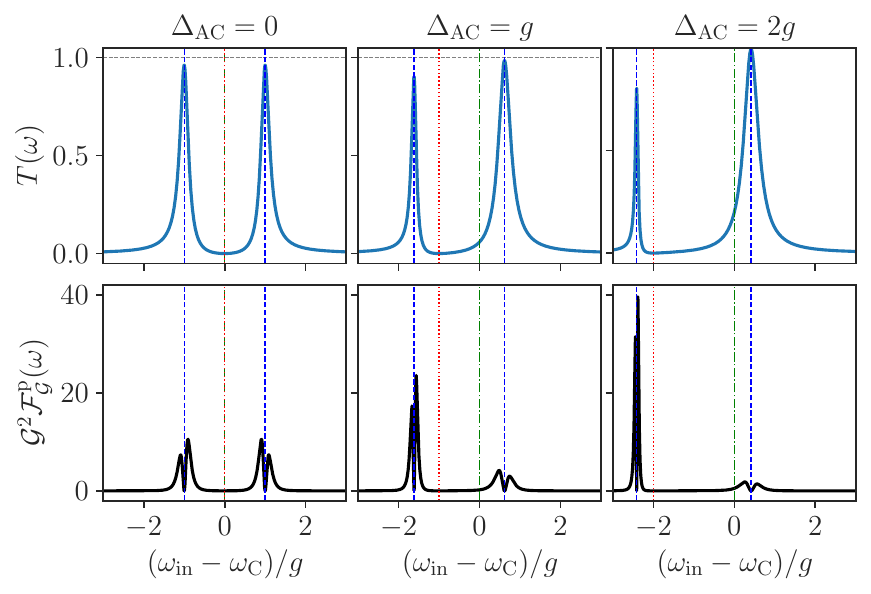} 
    \put(0,65){\textbf{(a)}}
    
    \end{overpic}    \begin{overpic}[scale=0.58]{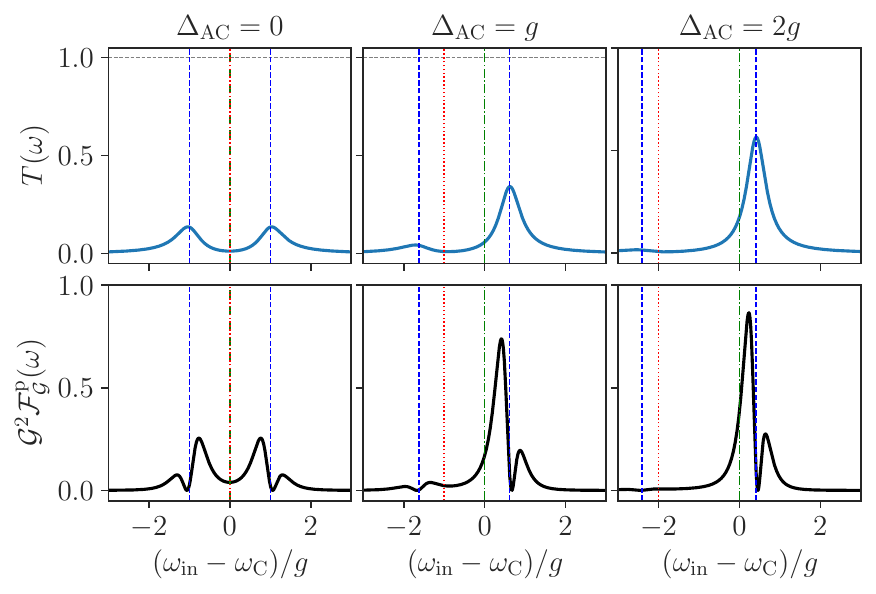} 
    \put(0,65){\textbf{(b)}}
    
    \end{overpic}
    
    \caption{ Transmission spectra $T \left( \omega \right)$(top panels) and the dimensionless FI (bottom panels) of photodetection for CEAS including atomic loss (a) $\Gamma=g/100$ and (b) $\Gamma=g$ are plotted against the detuning between the input probe frequency $\omega_{\mathrm{in}}$ and the cavity frequency, with solid blue and black lines respectively. The dashed blue lines represent the polariton frequencies $\epsilon_{\pm}$, while the green dash dotted lines and red dotted lines represent the bare frequencies of the cavity and atom respectively. We set $\mathcal{K}=g/2, \mathcal{K}_{\perp}=0, g=3.3\times 10^{-4} \mathrm{cm}^{-1}\approx 10 \mathrm{MHz}$ in these plots. 
  }
    \label{fig:CEAS_very_lossy}
\end{figure*}

\subsection{Semiclassical description of CES\label{SemiClassical}}

As shown in Fig.~\ref{fig:Losses_schematic} (a), the probe light -- CW and monochromatic with a fixed amplitude, illuminates a Fabry-Perot cavity containing the atom in a transmission geometry. This particular form of CES, where the output light is photo-detected, is called  cavity-enhanced absorption spectroscopy (CEAS). The transfer matrix method (TMM) relates the input and output fields on the two sides of a linear four-port optical device~\cite{Aoki2020}. Therein, the atom-cavity system is treated as a sequence of linear four-port devices for each of the elements - the mirrors (treated as beam-splitters), the atom and the free space between the atom and the cavity. The transfer matrices for these elements can be calculated independently and their product, in the order which they appear, is the transfer matrix of the total system. 

As the transfer matrix relates the fields on both sides,  the transmission function for the CEAS setup is (See Appendix~\ref{Transfer Matrix})

\begin{widetext}

\begin{align}
T\left(\omega_{\mathrm{in}}\right)=T\left(\Delta_{\mathrm{C}}-\omega_{\mathrm{C}}\right)=	\frac{\left|\mathcal{K}\left(\Gamma-2i\left(\Delta_{\mathrm{C}}+\Delta_{\mathrm{AC}}\right)\right)\right|^{2}}{\left|\mathcal{K}_{\mathrm{tot}}\left(\Gamma-2i\left(\Delta_{\mathrm{C}}+\Delta_{\mathrm{AC}}\right)\right)+4\left[\mathcal{G}-\Delta_{\mathrm{C}}\left(\Delta_{\mathrm{C}}+\Delta_{\mathrm{AC}}\right)\right]-2i\Delta_{\mathrm{C}}\Gamma\right|^{2}},
\label{eq:TransmissionFull}
\end{align}

\end{widetext}
where $\omega_{\mathrm{in}}$ is the input probe frequency and $\omega_{\mathrm{C}}$ is the frequency of the cavity mode, which has been assumed to be the fundamental mode of the Fabry-Perot cavity. $\Delta_{\mathrm{C}}=\omega_{\mathrm{in}}-\omega_{\mathrm{C}}$ and $\Delta_{\mathrm{AC}}=\omega_{\mathrm{C}}-\omega_{\mathrm{A}}$ represent the detunings between the probe-cavity and the atom-cavity frequencies respectively. $\Gamma$ is the rate of atomic loss to modes of the continuum that leak into the cavity and $\mathcal{K}$ is the rate of photon loss through the cavity mirrors. $\mathcal{K}_{\mathrm{tot}}=\mathcal{K}+\mathcal{K}_{\perp}$ is the total rate of photon loss through the cavity, where $\mathcal{K}_{\perp}$ is represents the rate of photon loss into the continuum due to imperfect transmission of light within the intra cavity free-space. The parameter which we seek to estimate is $\mathcal{G}=g^{2},$ the square of the coupling strength for the coherent interaction between the TLA and the cavity mode.

The output intensity, at a given probe frequency, is $I_{\mathrm{out}}\left(\omega_{\mathrm{in}}\right)=I_{0}T\left(\omega_{\mathrm{in}}\right)$. The average photon number that is detected by a photodetector with detection area $A_{\mathrm{d}}$ in a time interval $T_{\mathrm{d}}$ is $N_{\mathrm{out}}=I_{\mathrm{out}}\left(\omega_{\mathrm{in}}\right)A_{\mathrm{d}}T_{\mathrm{d}}/\hbar \omega_{\mathrm{in}}$. Due to the inevitable shot noise in the coherent CW drive, the measured output signal will be Poisson distributed about $N_{\mathrm{out}}$. The maximum attainable precision for estimating $\mathcal{G}$ in the presence of this shot noise is upper bounded by the FI for photon counting $\mathrm{FI}^{\mathrm{PC}}_{\mathcal{G}}= \left(\partial_{\mathcal{G}}N_{\mathrm{out}}\right)^{2}/N_{\mathrm{out}}$~\cite{FisherFlow}.

The FI per input photon, up to some factor determined by experimental parameters, can be written in terms of the transmission as~(See Appendix~\ref{Transfer Matrix})

\begin{align}
\mathcal{F}_{\mathcal{G}}^{\mathrm{p}}\left(\omega_{\mathrm{in}}\right)= & \frac{\left(\partial_{\mathcal{G}}T\left(\omega_{\mathrm{in}}\right)\right)^{2}}{T\left(\omega_{\mathrm{in}}\right)}.\label{eq:CES_FI_per_photon}
\end{align}

Fig.~\ref{fig:CEAS_very_lossy} shows the transmission and the dimensionless FI per photon $\mathcal{G}^2\mathcal{F}_{\mathcal{G}}^{\mathrm{p}}\left(\omega_{\mathrm{in}}\right)$ for two different values of the atomic loss $\Gamma$ and the atom-cavity detuning $\Delta_{\mathrm{AC}}$. See Appendix~\ref{TMM_lossy} for analytical expressions. The peaks in the transmission occur at the polariton frequencies $\left( \epsilon_{\pm}=\omega_{\mathrm{C}}+\frac{1}{2}\left(-\Delta_{\mathrm{AC}}\pm\Omega_{\mathrm{R}}\right)\right)$, which correspond to the dressed energy levels of the coupled atom–cavity system. These dressed modes, commonly referred to as polaritons, are hybrid light–matter states that arise in the strong coupling regime~\cite{larson2021jaynes}. Throughout this work, we take the value of the coupling strength to be $g\approx10 \mathrm{MHz}$, in accordance with achievable experimental values of couplings for atoms in optical cavities~\cite{Hood:98,Takahashi:20}.  All other frequencies have been scaled with respect to $g$. 

At low values of the atomic loss ($\Gamma=g/100$), near the polaritonic frequencies, almost all the light is transmitted, producing large transmission peaks which are symmetrically located when $\Delta_{\mathrm{AC}}=0$. However, as the detuning is increased, two effects become evident in the transmission. Firstly the polariton splitting becomes asymmetric as one of the peaks moves closer to the atomic frequency, becoming `atom-like', and the other moving closer to the cavity frequency, becoming `light-like' \cite{Mielke_1996,Rempe2005}. Additionally, the `atom-like'  peak becomes smaller due to the non-zero value of the atomic loss. Other than the broadening of the peaks, this general trend is still qualitatively present in the transmission peaks observed for large values of atomic loss $\Gamma=g$. 

\begin{figure*}[t]
    \centering
    \begin{minipage}{0.48\textwidth}
        \centering
        \includegraphics[width=\linewidth]{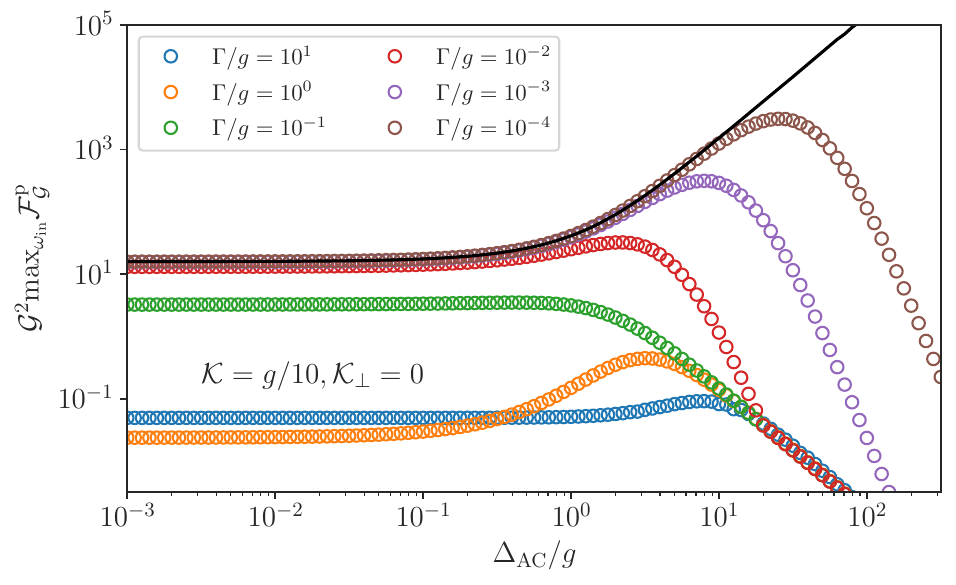}
        \caption{Dimensionless FI of photodetection for CEAS maximised over all input frequencies $\omega_{\mathrm{in}}$ is plotted against the atom-cavity detuning for different values of the atomic loss $\Gamma$. The monotonically increasing solid black line corresponds to $\Gamma=0.$}
        \label{fig:max_FI_CEAS_lossy}
    \end{minipage}
    \hfill
    \begin{minipage}{0.48\textwidth}
        \centering
        \includegraphics[width=\linewidth]{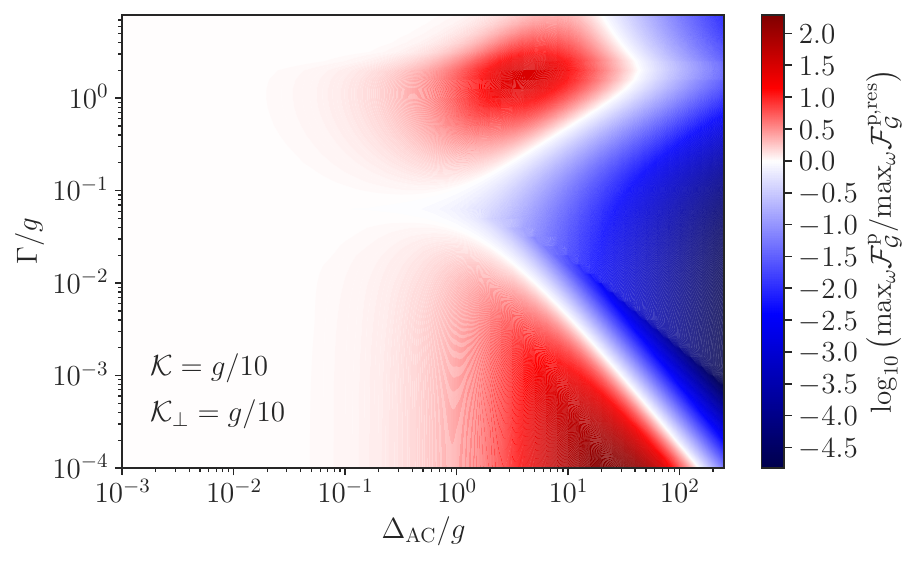}
        \caption{Dimensionless FI of photodetection for CEAS maximised over all input frequencies $\omega_{\mathrm{in}}$ is plotted against the atom-cavity detuning and the atomic loss. The red and blue shades correspond to regions in the parameter space where the FI is greater and smaller than the FI at resonance. }
    \label{fig:CEAS_CP_lossy}
    \end{minipage}
\end{figure*}

One of our main results is the behaviour of the FI peaks as the detuning increases and its qualitative difference for small and large $\Gamma.$ In the low loss regime, increasing detuning leads to higher FI coming from the polariton peak near the atomic frequency. This can be loosely understood as the `atom-like' polariton~\cite{Eberly} having more information about an atomic parameter. In the high loss regime, the `light-like' peak near the cavity frequency carries more FI as the detuning is increased. For these large values of the atomic loss, energy that reaches the atom (`atom-like') is predominantly lost to the environment and goes undetected. In contrast, the energy that primarily remains in the cavity (`light-like') survives the atomic loss, while still being sensitive to $\mathcal{G}$. Detuning the atom-cavity system suppresses the amount of excitation that reaches the atom. Consequently, the suppression of the energy exchange due to detuning results in the large FI values at the `light-like' peak. This is detuning-assisted enhancement (DAE).

Mathematically, the maximum value of the FI corresponds to the solutions of $\frac{d}{d\omega_{\mathrm{in}}}\mathcal{F}_{\mathcal{G}}^{\mathrm{p}}\left(\omega_{\mathrm{in}}\right)=0$. As, $\mathcal{F}_{\mathcal{G}}^{\mathrm{p}}\left(\omega_{\mathrm{in}}\right)$ is a ratio of complex fourth-order polynomials (see Appendix~\ref{TMM_lossy}), this amounts to solving for the roots of a seventh-order polynomial in $\omega_{\mathrm{in}}$ and finding which root maximises $\mathcal{F}_{\mathcal{G}}^{\mathrm{p}}$. This can be done numerically, providing the FI per photon maximised over the input probe frequency. This maximised FI is plotted against the atom-cavity detuning $\Delta_{\mathrm{AC}}$ for a range of $\Gamma$ values in Fig.~\ref{fig:max_FI_CEAS_lossy}. DAE is clearly observed across the weak ($\Gamma \gtrsim g$) and the strong coupling regimes ($\Gamma \ll g$). As $\Gamma$ becomes smaller, DAE is more pronounced and for zero atomic loss, the enhancement achievable through detuning becomes unbounded. This divergence is a property of the ideal lossless, infinitely narrowband model; in an experiment it will be regularized by finite probe linewidth of the laser drive.

The ratio of the maximised detuning-FI with the maximised at-resonance FI is also shown on a contour plot as a function of  $\Delta_{\mathrm{AC}}$ and $\Gamma$ in Fig.~\ref{fig:CEAS_CP_lossy}. The red zones in this plot correspond to a ratio greater than one, meaning that these are the regions where DAE is observed. Two distinct regions arise. The red region near the top, where we have large values of $\Gamma$, allows DAE because detuning acts as an inhibitor to the energy flowing into the atom and being lost to the environment. An intuitive explanation for the red region at the bottom, for small $\Gamma$ values, is that the FI for estimating an atomic parameter becomes large at the `atom-like' peaks of the transmission when atomic loss is small.

\subsection{CES with pulsed single-photon probes\label{Quantum}}

For a fuller understanding of DAE in this lossless regime, we now consider the case of CES with pulsed single-photon probes. While they carry starkly different photon statistics driving distinct cavity-atom dynamics, studying the fully quantum interaction of the TLA and the pulsed single-photon provides new insights into DAE in the absence of losses.

The quantum nature of the photon statistics of single-photon drive necessitates a quantum electrodynamical approach, which we will adopt here. The setup  consists of a TLA within a cavity mode which is interacting with a travelling light pulse (Fig.~\ref{fig:Losses_schematic} (b)). The atom-cavity  system reduces to a Jaynes-Cummings (JC) system under the assumption that the bare frequencies of both the atom as well as the cavity are much larger than the strength of their mutual coupling. The Hamiltonian for this system is
\begin{align}
    H_{\mathrm{ACP}}=&H_{\mathrm{JC}}+H_{\mathrm{P}}+H_{\mathrm{CP}},
    \label{eq:H_tot}
\end{align}
where $H_{\mathrm{JC}}$ represents the JC system containing the free atom and cavity terms along with the coherent interaction term, and $H_{\mathrm{P}}$ and $H_{\mathrm{CP}}$ correspond to the free Hamiltonian of the pulse modes and the interaction term between the cavity and the pulse. Note that the atom does not directly couple with the pulse and its interaction with the probe photon is mediated by the cavity. In the interaction picture with respect to the free Hamiltonians of the TLA, the cavity and the pulse modes, the Hamiltonian for the CES setup can be written as
\begin{align}
    H_{\mathrm{ACP}}^{\mathrm{Int}}\left(t\right)=&-i\hbar\left(ge^{-i\Delta_{\mathrm{AC}}t}\sigma^{\dagger}c+\sqrt{\mathcal{K}}ca^{\dagger}\left(t\right)-\mathrm{h.c}\right),
    \label{eq:H_int}
\end{align}
where $\sigma^{\dagger}$ represents the raising operator of the TLA, $c$ is the annihilation operator of the cavity mode, $\Delta_{\mathrm{AC}}=\omega_{\mathrm{C}}-\omega_{\mathrm{A}}$ is the atom-cavity detuning.
$a(t)$ is the so called `white-noise' operator of the pulse modes, introduced under the assumption that the spectral amplitude of the light pulse $\tilde{\xi}\left(\omega\right)$ is narrowband and slowly varying around some carrier frequency $\omega_{\mathrm{P}}$. It is defined as
\begin{align}
    a\left(t\right)=&\frac{e^{-i\Delta_{\mathrm{CP}}t}}{\sqrt{2\pi}}\int_{-\infty}^{+\infty}a\left(\Omega+\omega_{\mathrm{P}}\right)e^{-i\Omega t}d\Omega,
\end{align}
and satisfies $\left[a\left(t\right),a^{\dagger}\left(t'\right)\right]=\delta\left(t-t'\right)$. 

For pulsed single-photon probes, the initial state of the system is described by a temporal pulse shape $\xi\left(t\right)=\int_{-\infty}^{\infty}\frac{d\omega}{\sqrt{2\pi}}\tilde{\xi}\left(\omega\right)e^{i\left(\omega-\omega_{\mathrm{P}}\right)t}$, which is effectively an envelope that modulates oscillations at the carrier frequency and  is the Fourier transform of the spectral amplitude. In terms of the pulse shape, the initial state of the pulse is $\left|\psi_{\mathrm{P}}^{\mathrm{in}}\right\rangle =\int_{-\infty}^{+\infty}dt\xi\left(t\right)a^{\dagger}\left(t\right)\left|0\right\rangle $.

The number-conserving nature of Eq.~\eqref{eq:H_int} allows us to solve the corresponding Schr\"{o}dinger equation using an ansatz~\cite{Banacloche_2016,Francesco2023, Stolyarov:19,DarsheshdarMoniri2026}. See Appendix~\ref{Time evolution} for details. The JC system is initially in the ground state and is excited by the pulse. However, at asymptotically long times, the energy completely leaks out of the JC system and goes back into the pulse modes, meaning effectively that the information about the atomic parameter is encoded into the shape of the output pulse. The pulse at asymptotic times is a pure single-photon state and can be expressed as a unitary acting on the input state, i.e. $\left|\psi_{\mathrm{P}}^{\mathrm{out}}\right\rangle=U_{\mathrm{P}}\left|\psi_{\mathrm{P}}^{\mathrm{in}}\right\rangle$. This unitary is expressed in the frequency domain as  (see Eq.~\eqref{eq:Freq_domain_output_pulse})
\begin{align}
    U_{\mathrm{P}}&=\int_{-\infty}^{\infty}d\omega\exp\left\{ -i\varphi_{\mathcal{G}}\left(\omega\right)\right\} \left|1_{\omega}\right\rangle \left\langle 1_{\omega}\right|,
\end{align}
where 
\begin{align}
    \varphi_{\mathcal{G}}\left(\omega\right)=&2\tan^{-1}\left[\frac{\mathcal{K}\left(\omega+\Delta_{\mathrm{AC}}\right)}{2\left(\omega^{2}+\omega\Delta_{\mathrm{AC}}-\mathcal{G}\right)}\right]
    \label{Induced_phase}.
\end{align}

is the phase is imprinted on the output pulse by the interaction with the JC system, which encodes the information about the atomic parameter $\mathcal{G}$ into the output pulse.

Performing a measurement $\left\{\Pi_x=\left| x \right\rangle \left \langle x \right| \right\}$ on the output state yields a $\mathcal{G}$-dependent probability distribution $p_{\mathcal{G}}\left(x\right)=\text{Tr}\left(\rho_{\mathcal{G}}^{\mathrm{out}}\Pi_{x}\right)$, where $\rho_{\mathcal{G}}^{\mathrm{out}}=\left|\psi_{\mathrm{P}}^{\mathrm{out}}\right\rangle\left\langle\psi_{\mathrm{P}}^{\mathrm{out}}\right|$ is the density matrix corresponding to the output pulse. The measurement $\Pi_x$ projects the output state into the basis state $\left| x \right\rangle$ and $p_{\mathcal{G}}\left(x \right)$ is the probability of measuring  the output photon in that state. The maximum precision for estimating $\mathcal{G}$ from this distribution $p_{\mathcal{G}}\left(x\right)$ is the FI corresponding to the measurement $\Pi_x$. The FI is calculated from the probability distribution $p_{\mathcal{G}}(x)$ and is thus dependent on both the state $\rho_{\mathcal{G}}^{\mathrm{out}}$ and the measurement $\Pi_x$. The theoretical maximum of the FI over  all possible measurements on the quantum state $\left|\psi_{\mathrm{P}}^{\mathrm{out}}\right\rangle$ produces the quantum Fisher information (QFI) which is the ultimate precision that can be achieved for estimating $\mathcal{G}$ by measuring the output state  of the light~\cite{Helstrom:76,Holevo:11,Paris2009quantum}.

For unitary encodings of the parameter onto the input pulse, the QFI generated can be written as the variance of the phase sensitivity function $\mathcal{Z}_{\mathcal{G}}\left( \omega \right)=\partial_{\mathcal{G}}\varphi_{\mathcal{G}} \left( \omega \right)$ with respect to the input pulse $\left|\psi^{\mathrm{P}}_{\mathrm{out}}\right\rangle$~\cite{Liu:15,Braunstein:94,Braunstein:96,Pang2017Optimal}:
\begin{align}
    \mathcal{Q}_{\mathcal{G}}\left(\left|\psi_{\mathrm{in}}^{\mathrm{P}}\right\rangle \right)= & 4\left[\left\langle \psi^{\mathrm{in}}_{\mathrm{P}}\right|\mathcal{Z_{G}}^{2}\left|\psi^{\mathrm{in}}_{\mathrm{P}}\right\rangle -\left|\left\langle \psi^{\mathrm{in}}_{\mathrm{P}}\right|\mathcal{Z_{G}}\left|\psi^{\mathrm{in}}_{\mathrm{P}}\right\rangle \right|^{2}\right]\nonumber
    \\ 
    = & 4\mathrm{Var}\left(\mathcal{Z}_{\mathcal{G}}\right) 
    \label{eq:QFI},
\end{align}
where $\mathcal{Z}_{\mathcal{G}}$ is the $\mathcal{G}$-derivative of the operator that generates $U_{\mathrm{P}}$ and is expressed in the frequency basis as $\mathcal{Z}_{\mathcal{G}}=\int_{-\infty}^{\infty}d\omega\mathcal{Z}_{\mathcal{G}}\left(\omega\right)\left|1_{\omega}\right\rangle \left\langle 1_{\omega}\right|$. The optimal pulse shape that theoretically maximises the QFI is a coherent superposition of monochromatic frequencies $\tilde{\xi}_{\mathrm{opt}}\left(\omega\right)=2^{-\frac{1}{2}}\left[\delta\left(\omega-\omega_+\right)+\delta\left(\omega-\omega_-\right) \right]$, where $\omega_{\pm}$ are represent the global extremum points of  $\mathcal{Z}_{\mathcal{G}}\left(\omega\right)$~\cite{Pang2017Optimal,Pang_Brun_2014}. The QFI of this optimal pulse is a upper bound on the precision that can be obtained for the system and is written as
\begin{align}
    \mathcal{Q}_{\mathcal{G}}\leq \left[\mathcal{Z}_{\mathcal{G}}\left(\omega_+\right)-\mathcal{Z}_{\mathcal{G}}\left(\omega_-\right) \right]^2
    \label{QFI_bound_gen}.
\end{align}
Finding the extremum points of $\mathcal{Z}_{\mathcal{G}}\left(\omega\right)$ in general requires analytically solving for the roots of a quartic equation which is algebraically quite complicated. We thus find approximate extrema in two physically relevant regimes - the bad cavity (BC) $\mathcal{K}\gg \Omega_{\mathrm{R}}$, and the strong coupling (SC)  $\mathcal{K}\ll\Omega_{\mathrm{R}}$ regimes, where $\Omega_{\mathrm{R}}=\sqrt{\Delta_{\mathrm{AC}}^2+4\mathcal{G}}$ is the Rabi frequency of the JC system. The bounds on the dimensionless QFI in these limits are (See Appendix~\ref{QFI_bounds deriv}) 
\begin{align}
    \mathcal{G}^{2}\mathcal{\mathcal{Q}_{\mathcal{G}}^{\mathrm{BC}}}&\leq 4, \\
    \mathcal{G}^{2}\mathcal{Q}_{\mathcal{G}}^{\mathrm{SC}}\leq&16\frac{\Omega_{\mathrm{R}}^{2}}{\mathcal{K}^{2}}.
    \label{eq:SC_bounds}
\end{align}

In the BC regime, where the cavity decay is the fastest timescale, the cavity degree of freedom can be adiabatically eliminated by ignoring its fast dynamics. Hence the BC bound is identical to the QFI bound for pulsed single-photon spectroscopy in free-space~\cite{Sourav1}. In the SC regime, the bounds can be increased by decreasing the photon loss rate $\mathcal{K}$ through the mirrors, which corresponds to increasing the reflectivity of the cavity mirrors. The cavity-enhancement -- being proportional to the quality of the mirrors -- is also present in the semiclassical analysis, where the effective path length increases for better quality mirrors.

Our fully quantum analysis reveals a novel route to increasing the QFI bounds - by increasing the Rabi frequency. For a fixed value of the coupling $\mathcal{G}$, this can be achieved by increasing the atom-cavity detuning $\Delta_{\mathrm{AC}}$. 
This is DAE in a lossless CES setup, manifested analytically via the bounds on the QFI.

\begin{figure*}[t]
    \centering
    \begin{minipage}{0.48\textwidth}
        \centering
        \begin{overpic}
        [width=\columnwidth,percent]{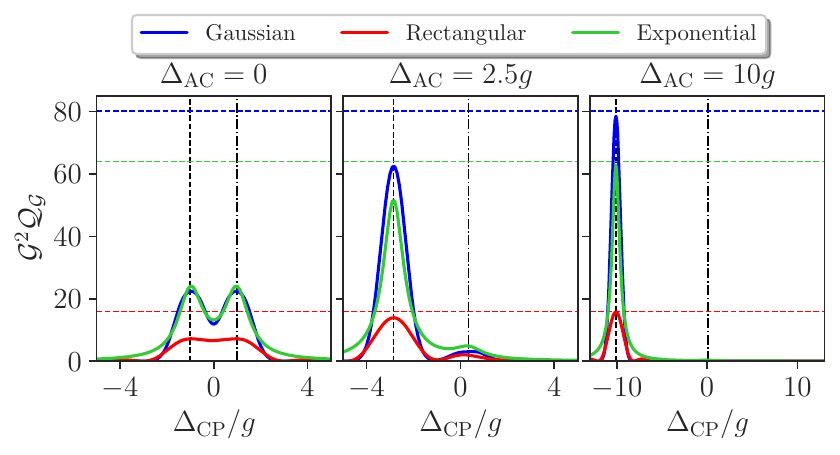}
        \put(33,30){\textbf{(a)}}
        \put(62,30){\textbf{(b)}}
        \put(91,30){\textbf{(c)}}
    \end{overpic}
    \caption{
    Dimensionless QFI for the single-photon pulses plotted against the pulse-cavity detuning $\Delta_{\mathrm{CP}}$ for different values of $\Delta_{\mathrm{AC}}/g$. The colors correspond to different temporal pulse shapes for the single-photon probe. The vertical black lines correspond to the two polariton frequencies. The horizontal dashed lines  represent the bounds on the QFI for the corresponding pulse shape, which are saturated at large detuning (Eq.~\eqref{eq:max_QFI_real}). We have set $\mathcal{K} = g/2$. $g=3.3\times 10^{-4} \mathrm{cm}^{-1}\approx 10 \mathrm{MHz}$ and $T=1/g$}. 
    \label{fig:DAE_SP}
    \end{minipage}
    \hfill
    \begin{minipage}{0.48\textwidth}
        \centering

        \begin{overpic}
        [width=0.98\columnwidth,percent]{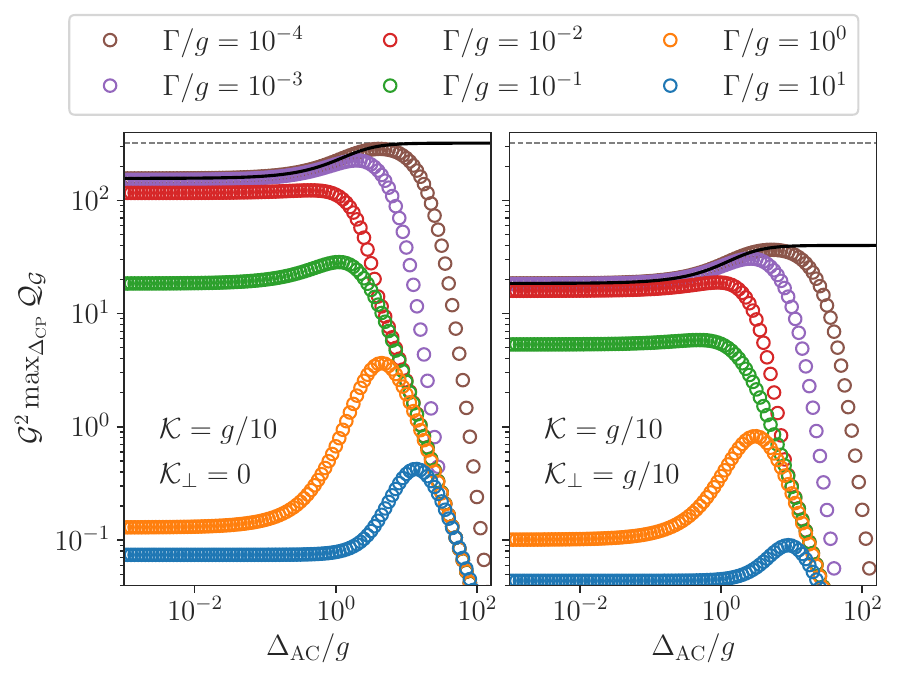}
        \put(17,47){\textbf{(a)}}
        \put(60,47){\textbf{(b)}}
        
    \end{overpic}
        \caption{Dimensionless QFI for the single-photon exponential pulse maximised over $\Delta_{\mathrm{CP}}$  is plotted against the atom-cavity detuning for different values of the atomic loss, represented by the different colors, for (a) $\mathcal{K}_{\perp}=0$ and (b) $\mathcal{K}_{\perp}=g/10$. The solid black line corresponds to the case with zero atomic loss. The dashed grey line represents the bound $\mathcal{Q}^{-}_{\mathcal{G},\mathrm{max}}$. The value of the JC coupling strength and the pulse duration were taken to be $g=3.3\times 10^{-4} \mathrm{cm}^{-1}\approx 10 \mathrm{MHz}$ and $T=1/g$ respectively.  Details of the QFI calculation in the presence of losses can be found in Appendix~\ref{SP_loss}.}
        \label{fig:DAE_QFI_lossy_main}
    \end{minipage}
\end{figure*}

Although the obtained bounds serve as a rigorous benchmark and describe the ideal configurations for obtaining enhanced precision, saturating these bounds requires the optimal pulse and measurement, which may be experimentally inaccessible. As an example of an experimentally realistic pulse shapes, we consider a exponentially rising input pulse. While  the white noise approximation sets the peak of the spectral amplitude $\tilde{\xi}\left(\omega \right)$ at the frequency of the cavity mode, we can `move' this peak by adding a shear to the pulse shape $\tilde{\xi}\left(\omega \right)\rightarrow\tilde{\xi}\left(\omega -\Delta_{\mathrm{CP}}\right)$, which detunes the carrier frequency of the pulse away from the cavity mode by $\Delta_{\mathrm{CP}}=\omega_{\mathrm{P}}-\omega_{\mathrm{C}}$. The output pulse shape can be analytically shown to have amplitudes which satisfy a dressed resonance condition (DRC) when the pulse peak hits polaritons frequencies $\left(\epsilon_{\pm} \right)$ of the JC system~(see Appendix~\ref{CFI_app}). This DRC produces large QFI values when the pulse shapes peak near the positions of the polariton frequencies, as shown in Fig.~\ref{fig:DAE_SP}(a).

In the SC regime where cavity-enhancement  is typically observed, the effect of atom-cavity detuning on the QFI for pulses peaked around the polariton frequencies can be analysed by approximating $\mathcal{Z}_{\mathcal{G}}\left( \omega \right)$  as a sum of delta functions at the polariton frequencies (See Eq.~\eqref{eq:Z_G_delta_approx}). Further details on how this approximation is performed are provided in Appendix~\ref{polariton_QFI_bounds}.  Decomposing $\mathcal{Z}_{\mathcal{G}}\left( \omega \right)$ allows us to obtain analytical expressions for the QFI of pulses that are peaked around the polariton frequencies using Eq.~\eqref{eq:QFI}. Using this approximate decomposition, QFI values for pulses peaked at the upper and lower polaritons (corresponding to the `light-like' and `atom-like' peaks respectively in the semiclassical picture) were found to be (See Eq.~\eqref{eq:polariton_bounds_QFI})
\begin{align}
    \mathcal{G}^{2}\mathcal{Q}_{\mathcal{G}}^{\pm}=&16\pi c_{\mathrm{P}}T\mathcal{G}^{2}\left[\frac{1}{2\mathcal{K}\mathcal{G}}\left(1\mp\frac{\Delta_{\mathrm{AC}}}{\Omega_{\mathrm{R}}}\right)-\frac{\pi c_{\mathrm{P}}T}{\Omega_{\mathrm{R}}^{2}}\right]\label{eq:DAE1}
\end{align}

where $T$ is the pulse duration of the temporal pulse shape $\xi\left(t\right)$ and $c_{\mathrm{P}}$ is a contant of order unity that depends on the pulse shape (See Eq.~\eqref{eq:Geometric_factor}). It is evident that setting $\Delta_{\mathrm{AC}}=0$ in the above equation results in $\mathcal{Q}_{\mathcal{G}}^{+}=\mathcal{Q}_{\mathcal{G}}^{-}$. Thus, at resonance, the QFI is equal for pulses centred at the two polariton frequencies as observed in Fig.~\ref{fig:DAE_SP}. At asymptotically large detunings ($\Delta_{\mathrm{AC}}^2\gg\mathcal{G}$), the ratio $\Delta_{\mathrm{AC}}/\Omega_{\mathrm{R}}$ goes to one. Thus, increasing the detuning leads to an increase in $\mathcal{Q}_{\mathcal{G}}^{-}$, which eventually saturates to 
\begin{equation}
\mathcal{Q}_{\mathcal{G},\mathrm{max}}^{-}=\frac{16\pi c_{\mathrm{P}}T}{\mathcal{K}\mathcal{G}}
\label{eq:max_QFI_real}
\end{equation}
as the last term in Eq. 12 vanishes in the large detuning limit. These are represented by the horizontal lines in Fig.~\ref{fig:DAE_SP}. In contrast, $\mathcal{Q}_{\mathcal{G}}^{+}$ vanishes as $\Delta_{\mathrm{AC}}$ becomes larger. This behaviour is clearly demonstrated by the  numerically computed QFI for non-zero $\Delta_{\mathrm{AC}}$ in Fig.~\ref{fig:DAE_SP}. Thus, Eq.~\eqref{eq:DAE1} clearly demonstrates that detuning leads to an enhancement of the maximum precision. We have also shown  (Fig.~\ref{fig:CFI_TCSPC}, Appendix~\ref{CFI_app})  that for pulses with long duration, experimentally accessible measurements like time correlated single-photon counting (TCSPC) can extract the DAE around the QFI peaks.  

We note that the bounds for the realistic pulse shapes in Eq.~\eqref{eq:DAE1} are always lower than the ones in Eq.~\eqref{eq:SC_bounds} which are for the optimal pulses and diverge at asymptotically large detuning. Although the maximum attainable QFI through realistic pulses at the negative polariton saturates to a finite value (Eq. \ref{eq:max_QFI_real}), this value diverges  in the limit of CW monochromatic light (where the pulse duration $T$ diverges), explaining the unbounded DAE that we also observed for the lossless case in Fig.~\ref{fig:max_FI_CEAS_lossy}.

\section{Discussion\label{Discussion}}

We identify an avenue to increase the precision of dipole moment estimation in CES by utilising the physics of cavity QED without significant additional experimental complexity.

Our results for the CW light demonstrate DAE to be achievable over a large range of values for the atomic loss when the cavity loss rates are smaller than the atom-cavity coupling rates. While the semiclassical description of CES for the CW probes is able to capture the DAE, the quantum description for pulsed single-photon probes allows us  to demonstrate the DAE as an effect that is also present at the level of the single-photons. As well as being observable in the QFI, the DAE should be  observable experimentally, as illustrated by the FI for typical single-photon measurements such as time-correlated single-photon counting(TCSPC).

DAE survives in the presence of atomic and cavity losses, and has the same qualitative behaviour as the CW light (See Fig.~\ref{fig:DAE_QFI_lossy_main}). The QFI maximised over the pulse-cavity detuning is higher at non-zero atom-cavity detuning compared to its values at resonance, but this effect does not survive at asymptotically large detunings or very large loss values. This is qualitatively similar to the behaviour of the CEAS FI in Fig.~\ref{fig:max_FI_CEAS_lossy}. However, in contrast to the semiclassical case, the maximised QFI in the absence of atomic losses saturates to a finite value at very high detuning and does not keep growing monotonically. For $\mathcal{K}_{\perp}=0$, the QFI saturates to the bounds derived for the lossless case in Eq.~\eqref{eq:DAE1}, whereas for non-zero cavity losses it saturates to some lower value. This hints at the existence of more general QFI bounds in the presence of losses, which could be explored in future work.   Additionally, we observe  that the large QFI peaks at non-zero detuning with low atomic loss are around the `atom-like' polariton, whereas for high atomic loss, they occur near the `light-like' polaritons (See Fig.~\ref{fig:lossy_QFI}), similar to the  behaviour in Fig.~\ref{fig:CEAS_CP_lossy} for CW probes.

While this work focuses on the estimation of dipole moments, it directly applies to estimation of other parameters from cavity transmission spectra, such as the number density of a gas when the dipole moment is known \cite{wright2023}. Indeed, our results open up the intriguing possibility that, more generally, other parameters inferred from cavity transmission spectra may be best estimated from off-resonance measurements. This challenges the intuitive assumption that the most informative measurement conditions involve zero detuning, as is conventionally the case in systems which exploit polaritonic effects \cite{thomas2024}. These potential benefits recommend a wider estimation theoretical analysis of spectroscopic measurements.

\paragraph{Acknowledgements:}

S.G. was supported by an EPSRC doctoral scholarship funded by the UKRI. Computing facilities were provided by the Scientific Research Technology Platform of the University of Warwick.

\bibliography{references}

\onecolumngrid
\appendix

\section{Local estimation theory \label{Est_Theory}}

Estimation theory provides a framework to quantify the precision of estimating a parameter $\theta$ from measurements of $x$ which are distributed according to a probability distribution $p_{\theta}\left(x\right)$, which has some parametric dependence on $\theta$. In quantum mechanics, this probability distribution is obtained from the collected data by the Born rule  $p_{\theta}\left(x\right)=\text{Tr}\left(\rho_{\theta}\Pi_{x}\right)$. Here $\rho_{\theta}$ is the quantum state which is dependent on the parameter $\theta$ and $\Pi_{x}$ is the positive operator valued measure(POVM) which mathematically describes the measurement whose possible experimental outcomes are labelled by $x$. For a photocounting measurement, $x$ would correspond to the number of photons detected. 

For an unbiased estimator of the parameter $\theta$, denoted by $\hat{\theta}$, its variance satisfies the Cram\'{e}r-Rao bound~\cite{Rao_1992, cramer2016}, which is given by

\begin{align}
\text{Var}\left[\hat{\theta}\right]\geq & \frac{1}{M\mathcal{C}\left(\rho_{\theta},\Pi_{x}\right)}\label{eq:Var_theta}
\end{align}

where $M$ denotes the number of repetitions of the measurement and $\mathcal{C}\left(\rho_{\theta},\Pi_{x}\right)$ is the classical Fisher information(CFI) for the estimation of $\theta$ from the state $\rho_{\theta}$ using the measurement described by $\Pi_{x}$. It is defined as

\begin{align}
\mathcal{C}\left(\rho_{\theta},\Pi_{x}\right)= & \sum_{x}\frac{1}{p_{\theta}\left(x\right)}\left(\frac{\partial p_{\theta}\left(x\right)}{\partial\theta}\right)^{2}
\end{align}

As Eq.~\eqref{eq:Var_theta} can be saturated with a large number of measurements by the maximum likelihood estimator, the CFI provides an upper bound on the maximum precision for estimating $\theta$ that can be achieved by the measurement $\Pi_{x}$. However to identify the quantum limit for the estimation of $\theta$, the CFI must be maximised over all possible measurements, obtaining the quantum Fisher information(QFI)~\cite{Paris2009quantum} $\mathcal{Q}\left(\rho_{\theta}\right)$, i.e

\begin{align}
\mathcal{Q}\left(\rho_{\theta}\right)= &\hspace{2pt} \underset{\left\{ \Pi_{x}\right\} }{\mathrm{max}}\hspace{2pt}\mathcal{C}\left(\rho_{\theta},\Pi_{x}\right)
\end{align}

The QFI is obtained mathematically as the expectation value of the square of the symmetric logarithmic derivative (SLD) operator $L_{\theta}$, which is an Hermitian operator satisfying the Lyapunov equation

\begin{align}
\frac{\partial\rho_{\theta}}{\partial\theta}= & \frac{\rho_{\theta}L_{\theta}+L_{\theta}\rho_{\theta}}{2}
\end{align}

In terms of the SLD, the QFI is expressed as

\begin{align}
\mathcal{Q}\left(\rho_{\theta}\right)= & \text{Tr}\left(\rho_{\theta}L_{\theta}^{2}\right),\label{eq:QFI_SLD}
\end{align}

which allows to write down the quantum Cram\'{e}r-Rao bound  on the variance

\begin{align}
\text{Var}\left[\hat{\theta}\right] & \geq\frac{1}{M\mathcal{C}\left(\rho_{\theta},\Pi_{x}\right)}\geq\frac{1}{M\mathcal{Q}\left(\rho_{\theta}\right)}
\end{align}

Under the assumption that $M$ is sufficiently large, CFI and QFI become the relevant metrics for the quantification of estimation precision. This formalism is known as local quantum estimation since the CFI and the QFI are calculated locally around the true value of the parameter $\theta$, meaning that $p_{\theta}$ and $\partial_{\theta}p_{\theta}$ are evaluated at the true value of the parameter in the equations above. It is also important to note that the QFI and the CFI are dimensional values when $\theta$ is not dimensionless. To allow comparison between different parameter values, we will define the dimensionless Fisher information values $\theta^{2}\mathcal{Q}\left(\rho_{\theta}\right)$ and $\theta^{2}\mathcal{C}\left(\rho_{\theta},\Pi_{x}\right)$. The dimensionless Fisher information sets an upper bound to the squared inverse coefficient of variation of the unbiased estimator $\theta^{2}/\text{Var}\left[\hat{\theta}\right]$, which is a dimensionless measure of the estimator's precision relative to the parameter being estimated. 

While it is not straightforward to put the expression for the QFI in Eq.~\eqref{eq:QFI_SLD} into an analytical closed form, and has to be evaluated by diagonalising the density matrix, it is possible to obtain simpler expressions for some special cases. For a pure state $\rho_{\theta}=\left|\Psi_{\theta}\right\rangle \left\langle \Psi_{\theta}\right|$, the QFI may be written as

\begin{align}
\mathcal{Q}\left(\left|\Psi_{\theta}\right\rangle \right)= & 4\left(\left\langle \partial_{\theta}\Psi_{\theta}\left|\partial_{\theta}\Psi_{\theta}\right.\right\rangle -\left|\left\langle \partial_{\theta}\Psi_{\theta}\left|\Psi_{\theta}\right.\right\rangle \right|^{2}\right)
\label{eq:QFI_pure}
\end{align}

Armed with the framework of local quantum estimation to obtain the precision with which we can estimate parameters from a quantum state, quantum light spectroscopy now boils down to a quantum optics problem where we have to model the interaction between the sample and the probe light and calculate the $\theta$-dependent output state of the light. 

In this section, we will calculate the QFI of the output pulses from the atom. By tracing out the atom and the cavity from $\left|\Psi^{\mathrm{ACF}}\left(t\right)\right\rangle \left\langle \Psi^{\mathrm{ACF}}\left(t\right)\right|$, we obtain the state of scattered light and write it (omitting the explicit time-dependence) as 
\begin{align}
\rho^{a}= & p^{\mathrm{JC}}\left|0^{a}\right\rangle \left\langle 0^{a}\right|+\left|\tilde{\psi}_{\mathrm{P}}\right\rangle \left\langle \tilde{\psi}_{\mathrm{P}}\right|
\end{align}

where $p^{\mathrm{JC}}$ is the probability of the excitation being inside the Jaynes-Cummings system, given by $p^{\mathrm{JC}}=\left|\psi_{\mathrm{A}}\right|^{2}+\left|\psi_{\mathrm{C}}\right|^{2}$ and $\left|\tilde{\psi}_{\mathrm{P}}\right\rangle $ is the unnormalised single-photon output pulse.

We are interested in the estimation of $\Gamma$, which can be written in terms of the atom-cavity coupling rate $g\left(\Gamma\right)$. Before we consider the QFI of $\Gamma$, we consider the QFI for an arbitrary parameter $\theta$ that is to be estimated from a given state. It is known that, for a general mixed state $\rho_{\theta}=\sum_{n}p_{n}\left|\psi_{n}\right\rangle \left\langle \psi_{n}\right|$, the QFI is given by

\begin{align}
\mathcal{Q}\left(\theta;\sum_{n}p_{n}\left|\psi_{n}\right\rangle \left\langle \psi_{n}\right|\right)= & \sum_{n}\frac{\left(\partial_{\theta}p_{n}\right)^{2}}{p_{n}}+\sum_{n}4p_{n}\left\langle \partial_{\theta}\psi_{n}\left|\partial_{\theta}\psi_{n}\right.\right\rangle \\
 & -\sum_{m,n}\frac{8p_{m}p_{n}}{p_{m}+p_{n}}\left|\left\langle \partial_{\theta}\psi_{m}\left|\psi_{n}\right.\right\rangle \right|^{2}\nonumber 
\end{align}

Thus, applying this expression for the light states of the form obtained above, i.e., $\rho_{\theta}=p_{\theta}\left|0\right\rangle \left\langle 0\right|+\left(1-p_{\theta}\right)\left|\psi_{\theta}\right\rangle \left\langle \psi_{\theta}\right|$, we get

\begin{align}
\mathcal{Q}\left(\theta;\rho\right)= & \frac{\left(\partial_{\theta}p_{\theta}\right)^{2}}{p_{\theta}\left(1-p_{\theta}\right)}+4\left(1-p_{\theta}\right)\left[\left\langle \partial_{\theta}\psi_{\theta}\left|\partial_{\theta}\psi_{\theta}\right.\right\rangle -\left|\left\langle \partial_{\theta}\psi_{\theta}\left|\psi_{\theta}\right.\right\rangle \right|^{2}\right]\nonumber \\
= & \mathcal{C}\left(p_{\theta}\right)+\left(1-p_{\theta}\right)\mathcal{Q}\left(\left|\psi_{\theta}\right\rangle \right)\label{eq:QFI_Gen_out}
\end{align}

where $\mathcal{C}\left(p_{\theta}\right)$ is the classical Fisher information associated with a distribution given by $\left[p_{\theta},1-p_{\theta}\right]$ and $\mathcal{Q}\left(\left|\psi_{\theta}\right\rangle \right)$ is the QFI of the pure single-photon state. Under the assumption that all the amplitudes are real and the above expression can be written directly in terms of the unnormalised single-photon states $\left|\tilde{\psi}_{\theta}\right\rangle =\sqrt{1-p_{\theta}}\left|\psi_{\theta}\right\rangle $ as follows

\begin{align}
\mathcal{Q}\left(\theta;\rho_{\theta}\right)= & \frac{\left(\partial_{\theta}p_{\theta}\right)^{2}}{p_{\theta}}+4\left\langle \partial_{\theta}\tilde{\psi}_{\theta}\left|\partial_{\theta}\tilde{\psi}_{\theta}\right.\right\rangle \label{eq:QFI_gen_out}
\end{align}

We can always write down the QFI of another parameter which is given by $\chi=\chi\left(\theta\right)$. To do this, we can simply relabel the state as $\rho_{\theta}\rightarrow\rho_{\chi\left(\theta\right)}$ and note that 
\begin{align}
\mathcal{Q}\left(\chi;\rho_{\chi\left(\theta\right)}\right)= & \frac{\left(\partial_{\chi}p_{\chi}\right)^{2}}{p_{\chi}}+4\left\langle \partial_{\chi}\tilde{\psi}_{\chi}\left|\partial_{\chi}\tilde{\psi}_{\chi}\right.\right\rangle 
\end{align}

Now, with respect to to this state $\rho_{\chi\left(\theta\right)}$, we can also write down the QFI of the original parameter $\theta$. Noting that $\partial_{\theta}\left(\dots\right)=\left(\partial_{\theta}\chi\right)\partial_{\chi}\left(\dots\right)$ this is given by

\begin{align}
\mathcal{Q}\left(\theta;\rho_{\chi\left(\theta\right)}\right)= & \left(\partial_{\theta}\chi\right)^{2}\frac{\left(\partial_{\chi}p_{\chi\left(\theta\right)}\right)^{2}}{p_{\chi}}+4\left(\partial_{\theta}\chi\right)^{2}\left\langle \partial_{\chi}\tilde{\psi}_{\chi\left(\theta\right)}\left|\partial_{\chi}\tilde{\psi}_{\chi\left(\theta\right)}\right.\right\rangle \nonumber \\
= & \left(\partial_{\theta}\chi\right)^{2}\mathcal{Q}\left(\chi;\rho_{\chi\left(\theta\right)}\right)\label{eq:QFI_rel}
\end{align}

Therefore, for a given state, the relation of the QFI for two parameters can be related by Eq.~\eqref{eq:QFI_rel} as long as the two parameters have a functional relation that is known. Although the above relation was derived here under the assumption of real probability amplitudes, it is easy to see that it holds for the general QFI expression of the scattered output light in Eq.~\eqref{eq:QFI_Gen_out}. If the functional relation between $\theta$ and $\chi$ is linear, i.e. if $\chi=\lambda\theta$, the derivative is given by $\partial_{\theta}\chi=\lambda$. In this case, if we multiply both sides of Eq.~\eqref{eq:QFI_rel} by $\theta^{2}$, we see that the dimensionless QFI for both the parameters are exactly equal

\begin{align}
\theta^{2}\mathcal{Q}\left(\theta;\rho_{\chi\left(\theta\right)}\right)= & \chi^{2}\mathcal{Q}\left(\chi;\rho_{\chi\left(\theta\right)}\right), \hspace{1.0cm} \mathrm{if}  \hspace{5pt}\chi=\lambda\theta\label{eq:QFI_lin_rel}
\end{align}

As our aim is to probe the atom by measurements of the output light, the parameter of interest to us is the dipole moment of the atom. In our case, the Jaynes-Cummings atom-cavity coupling $g$ is directly proportional to the dipole moment of the atom along the polarisation of the cavity field. In previous works on quantum light spectroscopy, the free space atom-field coupling $\Gamma$, which is proportional to the linewidth of the atom, was estimated. The relation between $g$ and $\Gamma$ is given by

\begin{align}
g\left(\Gamma\right) & =\sqrt{\lambda\left(\omega_{\mathrm{C}},\omega_{\mathrm{A}},\omega_{\mathrm{P}}\right)\Gamma}
\end{align}

We define a new parameter $\mathcal{G}=g^{2}$, allowing us to obtain the linear relationship with $\Gamma$. This gives us

\begin{align}
\mathcal{G}= & \lambda\Gamma
\end{align}
where $\lambda$ has been taken to be a constant that is not dependent on the dipole moment. The above linear scaling ensures that the dimensionless QFI that we obtain for the estimation of $\mathcal{G}$ is always equal to the dimensionless QFI for estimation of $\Gamma$. Thus, for a given output pulse state, we have

\begin{align}
\mathcal{G}^{2}\mathcal{Q}\left(\mathcal{G};\rho_{\mathcal{G}}^{\mathrm{out}}\right)= & \Gamma^{2}\mathcal{Q}\left(\Gamma;\rho^{\mathrm{out}}\right)
\end{align}

Since the expressions of the output pulse obtained after scattering with the cavity contain the parameter $g\sim\sqrt{\mathcal{G}}$, we re-write the output pulse labelled as
\begin{align}
\rho_{\mathcal{G}}^{\mathrm{out}}= & p_{\mathcal{G}}\left|0^{a}\right\rangle \left\langle 0^{a}\right|+\left(1-p_{\mathcal{G}}\right)\left|\psi_{\mathrm{P}}\right\rangle \left\langle \psi_{\mathrm{P}}\right|
\end{align}
where $p_{\mathcal{G}}$ is the $\mathcal{G}$-dependent probability of the excitation being in the JC system, therefore $1-p_{\mathcal{G}}=\left\langle \tilde{\psi}_{\mathrm{P}}\left|\tilde{\psi}_{\mathrm{P}}\right.\right\rangle $ and $\left|\psi_{\mathrm{P}}\right\rangle =\frac{\left|\tilde{\psi}_{\mathrm{P}}\right\rangle }{\sqrt{\langle \tilde{\psi}_{\mathrm{P}}\left|\tilde{\psi}_{\mathrm{P}}\right\rangle} }$. We can also compute the QFI for the estimation of the detuning of the JC system, given by $\Delta_{\mathrm{AC}}$. Under the assumption that we know the frequency of the cavity, estimating the detuning is equivalent to estimating the transition frequency of the atomic degree of freedom. In the following sections, we will consider the estimation of both these parameters, which together characterize the atom completely.

\section{Transfer matrix description of an atom-cavity system\label{Transfer Matrix}}

The typical implementation of cavity-enhanced absorption spectroscopy(CEAS) depends on how the atom inside the optical cavity modifies the resonant transmission of light through a cavity. This modified transmission can be calculated by a transfer matrix analysis by treating the atom-cavity system as a linear four-port device. The input and output fields on the left and right sides of the cavity mirrors can are related by the following expression

\begin{align}
\left(\begin{array}{c}
E_{\mathrm{L}}^{\left(\mathrm{in}\right)}\\
E_{\mathrm{L}}^{\left(\mathrm{out}\right)}
\end{array}\right)= & \mathcal{T}_{\mathrm{AC}}\left(\begin{array}{c}
E_{\mathrm{R}}^{\left(\mathrm{out}\right)}\\
E_{\mathrm{R}}^{\left(\mathrm{in}\right)}
\end{array}\right)=\left(\begin{array}{c}
\mathcal{T}_{11}E_{\mathrm{R}}^{\left(\mathrm{out}\right)}+\mathcal{T}_{12}E_{\mathrm{R}}^{\left(\mathrm{in}\right)}\\
\mathcal{T}_{21}E_{\mathrm{R}}^{\left(\mathrm{out}\right)}+\mathcal{T}_{22}E_{\mathrm{R}}^{\left(\mathrm{in}\right)}
\end{array}\right)
\end{align}

where $\mathcal{T}_{ij}$ are the elements of the transfer matrix for the atom-cavity system denoted as $\mathcal{T}_{\mathrm{AC}}$. Under the assumption that there is no input light from the right hand side, the output light on the right hand side is given by

\begin{align}
E_{\mathrm{R}}^{\left(\mathrm{out}\right)}= & \frac{1}{\mathcal{T}_{11}}E_{\mathrm{L}}^{\left(\mathrm{in}\right)}
\end{align}

In the limit of a high finesse cavity with no atomic loss and under the assumption that the detuning between the probe frequency and cavity mode(which is assumed to be the fundamental mode) is small, the transmission through the atom-cavity system is written as a function of the input probe frequency as

\begin{align}
T\left(\omega_{\mathrm{in}}\right)= & \left|\frac{1}{\mathcal{T}_{11}}\right|^{2}=\left|\frac{-i\frac{\mathcal{K}}{2}\left(\omega_{\mathrm{in}}-\omega_{\mathrm{A}}\right)}{-i\left[\frac{\mathcal{K}}{2}-i\left(\omega_{\mathrm{in}}-\omega_{\mathrm{C}}\right)\right]\left(\omega_{\mathrm{in}}-\omega_{\mathrm{A}}\right)+\mathcal{G}}\right|^{2}
\end{align}

Introducing $\Delta_{\mathrm{A/C}}=\omega_{\mathrm{in}}-\omega_{\mathrm{A/C}}$ and using the atom-cavity detuning $\Delta_{\mathrm{CA}}=\omega_{\mathrm{A}}-\omega_{\mathrm{C}}$, the above expression can be recast as

\begin{align}
T\left(\Delta_{\mathrm{C}}+\omega_{\mathrm{C}}\right)= & \sin^{2}\left(\frac{\varphi\left(\Delta_{\mathrm{C}}\right)}{2}\right)
\end{align}

This $\varphi$ is the same phase that is induced in the output pulse for the single-photon light. It can be shown that in the strong coupling limit, for frequencies resonant with the poles of the system (which are the eigen-energies of the JC system) this phase is given by $\varphi\left(\epsilon_{\pm}\right)=\pi$. When the input probe frequency matches these polariton frequencies, the transmission becomes one, meaning that the entire input light is entirely transmitted at these frequencies. 

When the input field intensity is given by $I_{0}$, the output field intensity, for probe frequencies detuned from the cavity by $\Delta_{\mathrm{C}}$, is given by

\begin{align}
I_{\mathrm{out}}\left(\omega_{\mathrm{in}}\right)= & I_{0}\sin^{2}\left(\frac{\varphi\left(\Delta_{\mathrm{C}}\right)}{2}\right)
\end{align}

Thus the number of photons detected by a detector with area $A_{\mathrm{det}}$, within a time interval $T_{\mathrm{det}}$ is given by

\begin{align}
N_{\mathrm{out}}\left(\omega_{\mathrm{in}}\right)= & \frac{I_{\mathrm{out}}\left(\omega_{\mathrm{in}}\right)}{\hbar\omega_{\mathrm{in}}}A_{\mathrm{det}}T_{\mathrm{det}}
\end{align}

In any actual spectroscopic measurement, the precision of photo-counting will be limited by the shot noise, which means the output signal will be Poisson distributed about the mean number of photons. The Fisher information for the measurement of a parameter $\theta$ may then be written as

\begin{align}
\mathrm{FI}^{\mathrm{PC}}_{\theta}= & \frac{\left(\partial_{\mathcal{\theta}}N_{\mathrm{out}}\right)^{2}}{N_{\mathrm{out}}}
\end{align}

where the PC super-script indicates that this is the FI of photo-counting. Since $N_{\mathrm{out}}=N_{\mathrm{out}}\left(\omega_{\mathrm{in}}\right)$, the Fisher information for CEAS is given as a function of the input probe frequency, as every different input frequency corresponds to a particular spectroscopic setup and repeated measurements at this input frequency will yield a Poisson distribution due to shot noise. The Fisher information for estimation of $\mathcal{G}$ per unit detection time and detector area, can be  written down analytically in terms of the phase $\varphi$ as

\begin{align}
\mathcal{F_{G}}\left(\omega_{\mathrm{in}}\right)=\frac{\mathrm{FI}^{\mathrm{PC}}_{\mathcal{G}}}{A_{\mathrm{det}}T_{\mathrm{det}}}= & \frac{I_{0}}{\hbar\omega_{\mathrm{in}}}\left[\cos\left(\frac{\varphi\left(\Delta_{\mathrm{C}}\right)}{2}\right)\mathcal{Z}_{\mathcal{G}}\left(\Delta_{\mathrm{C}}\right)\right]^{2}
\end{align}

where $\mathcal{Z}_{\mathcal{G}}=\partial_{\mathcal{G}}\varphi$ is the phase sensitivity function written down previously. Meanwhile, the total Fisher information for the estimation of $\mathcal{G}$ will scale with the detector area and time of detection. It is seen from the plots that higher precision is achieved when the probe frequency $\omega_{\mathrm{in}}$ is near the poles of the JC system. This is expected as the transmission peaks appear when the probe frequency nears these poles, producing a larger photon output photon flux and thereby lowering shot noise. Both these peaks have an equivalent amount of information about the parameter $\mathcal{G}$ when the atom and the cavity are at resonance. However as the detuning between the atom and the cavity is increased, more information is carried by the transmission peak at the negative pole, which is equivalent to the fluorescense peak of the atom. Moreover, the Fisher information carried by these peaks are much larger than Fisher information carried by the twin peaks found at resonance. Thus, we may conclude that for the lossless case, detuning between the atom and the cavity monotonically increases the maximum Fisher information or the maximum precision that maybe achieved in CEAS. 

It is also useful to look at another quantity, which is the Fisher information for $\mathcal{G}$ estimation per input photon, which is given by

\begin{align}
\mathcal{F}_{\mathcal{G}}^{\mathrm{p}}\left(\omega_{\mathrm{in}}\right)= & \frac{\mathrm{FI}^{\mathrm{PC}}_{\mathcal{G}}}{N_{\mathrm{in}}}
\end{align}

where $N_{\mathrm{in}}$ represents the number of input photons, given by 
\begin{equation}
N_{\mathrm{in}}\left(\omega_{\mathrm{in}}\right)=\frac{I_{0}}{\hbar\omega_{\mathrm{in}}}A_{\mathrm{in}}T_{\mathrm{in}}
\end{equation}

where $A_{\mathrm{in}}$ and $T_{\mathrm{in}}$ are the cross-sectional area of the input laser beam and the duration for which it is kept on. Using the expressions, derived previously, it can be shown that

\begin{align}
\mathcal{F}_{\mathcal{G}}^{\mathrm{p}}\left(\omega_{\mathrm{in}}\right)= & \frac{\frac{I_{0}}{\hbar\omega_{\mathrm{in}}}\left[\cos\left(\frac{\varphi\left(\Delta_{\mathrm{C}}\right)}{2}\right)\mathcal{Z}_{\mathcal{G}}\left(\Delta_{\mathrm{C}}\right)\right]^{2}A_{\mathrm{det}}T_{\mathrm{det}}}{\frac{I_{0}}{\hbar\omega_{\mathrm{in}}}A_{\mathrm{in}}T_{\mathrm{in}}}\nonumber \\
= & \eta\left[\cos\left(\frac{\varphi\left(\Delta_{\mathrm{C}}\right)}{2}\right)\mathcal{Z}_{\mathcal{G}}\left(\Delta_{\mathrm{C}}\right)\right]^{2},
\end{align}
where $\eta=\frac{A_{\mathrm{det}}T_{\mathrm{det}}}{A_{\mathrm{in}}T_{\mathrm{in}}}$ is a dimensionless quantity that is determined by the experimental parameters and can be set to unity. 

\subsection{Non-zero losses\label{TMM_lossy}}

The above calculations have assumed the absence of any losses in the spectroscopic setup. However, there are broadly two types of losses that can be present in the CEAS setup - the loss from the atomic TLA due to dephasing or decay into the continuum modes that leak into the cavity and the loss directly from the cavity which occurs when the linear transmission between the cavity mirrors is not perfectly efficient. Under the presence of these losses, the transmission is modified as

\begin{align}
T\left(\omega_{\mathrm{in}}\right)=T\left(\Delta_{\mathrm{C}}+\omega_{\mathrm{C}}\right)= & \frac{\left|\mathcal{K}\left(\Gamma-2i\left(\Delta_{\mathrm{C}}+\Delta_{\mathrm{AC}}\right)\right)\right|^{2}}{\left|\mathcal{K}_{\mathrm{tot}}\left(\Gamma-2i\left(\Delta_{\mathrm{C}}+\Delta_{\mathrm{AC}}\right)\right)+4\left[\mathcal{G}-\Delta_{\mathrm{C}}\left(\Delta_{\mathrm{C}}+\Delta_{\mathrm{AC}}\right)\right]-2i\Delta_{\mathrm{C}}\Gamma\right|^{2}}
\end{align}
where we have $\mathcal{K}_{\mathrm{tot}}=\mathcal{K}+\mathcal{K}_{\perp}$ and $\Gamma,\mathcal{K}_{\perp}$ represent the losses from the atom and the cavity respectively. This modified transmission function in the presence of losses will lead to a difference in the mean number of transmitted photons, about which the output signal is Poisson distributed. Thus the inclusion of these losses will lead to changes in the Fisher information as well. The Fisher information for estimation of $\mathcal{G}$ per unit detection time and detector area, in the presence of these losses can be written as

\begin{align}
\mathcal{F_{G}}\left(\omega_{\mathrm{in}}\right)=\frac{\mathrm{FI}_{\mathrm{CEAS}}}{A_{\mathrm{det}}T_{\mathrm{det}}}= & \frac{I_{0}}{\hbar\omega_{\mathrm{in}}}T\left(\omega_{\mathrm{in}}\right)\left[\partial_{\mathcal{G}}\left(\ln f_{\mathcal{G}}\left(\Delta_{\mathrm{C}}\right)\right)\right]^{2}
\end{align}
where $f_{\mathcal{G}}\left(\Delta_{\mathrm{C}}\right)$ is the denominator of the transmission function, given by

\begin{align}
f_{\mathcal{G}}\left(\Delta_{\mathrm{C}}\right)= & \left|\mathcal{K}_{\mathrm{tot}}\left(\Gamma-2i\left(\Delta_{\mathrm{C}}+\Delta_{\mathrm{AC}}\right)\right)+4\left[\mathcal{G}-\Delta_{\mathrm{C}}\left(\Delta_{\mathrm{C}}+\Delta_{\mathrm{AC}}\right)\right]-2i\Delta_{\mathrm{C}}\Gamma\right|^{2}
\end{align}

The Fisher information per input photon is given by

\begin{align}
\mathcal{F}_{\mathcal{G}}^{\mathrm{p}}\left(\omega_{\mathrm{in}}\right)= & \frac{\mathrm{FI}_{\mathrm{CEAS}}}{N_{\mathrm{in}}}=\frac{\left(\partial_{\mathcal{\theta}}N_{\mathrm{out}}\right)^{2}}{N_{\mathrm{out}}N_{\mathrm{in}}}\nonumber \\
= & \eta \frac{\left[\partial_{\mathcal{G}}T\left(\omega_{\mathrm{in}}\right)\right]^{2}}{T\left(\omega_{\mathrm{in}}\right)}
\end{align}

where $\eta$ is the same dimensionless experimental parameter and can be set to unity. The optimal input frequency can be found by solving for the extremum points of $\mathcal{F}_{\mathcal{G}}^{\mathrm{p}}\left(\omega_{\mathrm{in}}\right)$, found at the roots of $\partial_{\omega_{\mathrm{in}}}\mathcal{F}_{\mathcal{G}}^{\mathrm{p}}\left(\omega_{\mathrm{in}}\right)=0$. Due to the form of $T\left(\omega_{\mathrm{in}}\right)$, finding the optimal input is equivalent to calculating the roots of a seventh-order polynomial in $\omega_{\mathrm{in}}$ which is complicated analytically. In practice, the optimal frequencies can be numerically obtained by numerical root solvers. In the regimes where the cavity loss rate is small compared to the atom-cavity coupling, the optimal frequencies are found near the polariton frequencies. At resonance, the FI at both the frequencies are equal, however for the detuned system, the optimal polariton depends on the atomic loss. At large detuning, the optimal frequency is found near the `light-like' polariton when the atomic loss is high and it is found near the `atom-like' polariton when the atomic loss is small compared to the atom-cavity coupling rate.

\section{Time evolution of the JC system under pulsed single-photon drive\label{Time evolution}}

We will consider the case where the Jaynes-Cummings(JC) system formed by the TLA and the cavity is initially in its ground state $\left|g,0^{c}\right\rangle $ and a single-photon pulse $\left|\tilde{\psi}_{\mathrm{P}}^{a}\left(t_{0}\right)\right\rangle =\int_{-\infty}^{+\infty}d\tau\xi\left(\tau\right)a^{\dag}\left(\tau\right)\left|0^{a}\right\rangle $ is sent in via the light channel. The absence of any non-number-conserving terms in the interaction Hamiltonian, ensures that the system will remain in the single excitation subspace at all times. This allows us to write an ansatz of the following form

\begin{align}
\left|\Psi^{\mathrm{ACF}}\left(t\right)\right\rangle = & \psi_{\mathrm{A}}\left(t\right)\left|e\right\rangle \left|0^{c}\right\rangle \left|0^{a}\right\rangle +\psi_{\mathrm{C}}\left(t\right)\left|g\right\rangle \left|1^{c}\right\rangle \left|0^{a}\right\rangle \label{eq:Ansatz}\\
 & +\left|g\right\rangle \left|0^{c}\right\rangle \left|\tilde{\psi}_{\mathrm{P}}\left(t\right)\right\rangle \nonumber 
\end{align}
where $\psi_{\mathrm{A}}$ is the atom excitation amplitude (AEA), $\psi_{\mathrm{C}}$ is the probability amplitude for the cavity being in its first excited state, henceforth referred to as the cavity excitation amplitude (CEA) and $\left|\tilde{\psi}_{\mathrm{P}}\left(t\right)\right\rangle $ is the unnormalised single-photon state in the light interacting with the cavity. Applying this ansatz to the time-dependent Schr\"{o}dinger equation for the Hamiltonian in Eq.~\eqref{eq:H_int}, we have $i\hbar\partial_{t}\left|\Psi^{\mathrm{ACF}}\left(t\right)\right\rangle =H_{\mathrm{I}}^{\mathrm{ACF}}\left(t\right)\left|\Psi^{\mathrm{ACF}}\left(t\right)\right\rangle $. Following the prescription described in~\cite{Banacloche_2016} we obtain the following coupled dynamical equations for the time evolution of the unknown values 

\begin{align}
\psi_{\mathrm{A}}\left(t\right)= & -g\int_{-\infty}^{t}dt_{1}e^{-i\Delta_{\mathrm{AC}}t_{1}}\psi_{\mathrm{C}}\left(t_{1}\right)\\
\psi_{\mathrm{C}}\left(t\right)= & g\int_{-\infty}^{t}dt_{1}e^{+i\Delta_{\mathrm{AC}}t_{1}}e^{-\frac{\mathcal{K}}{2}\left(t-t_{1}\right)}\psi_{\mathrm{A}}\left(t_{1}\right)+\sqrt{\mathcal{K}}\int_{-\infty}^{t}dt_{1}e^{-\frac{\mathcal{K}}{2}\left(t-t_{1}\right)}\xi\left(t_{1}\right)\\
\left|\tilde{\psi}_{\mathrm{P}}\left(t\right)\right\rangle = & \int_{-\infty}^{+\infty}d\tau\tilde{\psi}_{\mathrm{P}}\left(t,\tau\right)a^{\dag}\left(\tau\right)\left|0^{a}\right\rangle \label{eq:Out_light_state}
\end{align}

where 
\begin{align}
\tilde{\psi}_{\mathrm{P}}\left(t,\tau\right)= & \xi\left(\tau\right)-\sqrt{\mathcal{K}}\Theta\left(t-\tau\right)\psi_{\mathrm{C}}\left(\tau\right)\label{out_light_shape}
\end{align}

To decouple the equations for the atomic and cavity excitation amplitudes we will substitute the expression for $\psi_{\mathrm{A}}\left(t\right)$ into the above expression for $\psi_{\mathrm{C}}\left(t\right)$, which allows us to write

\begin{align}
\psi_{\mathrm{C}}\left(t\right)= & -g^{2}\int_{-\infty}^{t}dt_{1}\int_{-\infty}^{t_{1}}dt_{2}e^{+i\Delta_{\mathrm{AC}}\left(t_{1}-t_{2}\right)}e^{-\frac{\mathcal{K}}{2}\left(t-t_{1}\right)}\psi_{\mathrm{C}}\left(t_{2}\right)+\sqrt{\mathcal{K}}\int_{-\infty}^{t}dt_{1}e^{-\frac{\mathcal{K}}{2}\left(t-t_{1}\right)}\xi\left(t_{1}\right)\label{eq:cav_dyn2}
\end{align}

At this point, we note that setting $g=0$ which corresponds to removing the atom from the cavity, allows us to recover the expressions for the excitation amplitudes derived in~\cite{Francesco2023} for a TLA interacting with a single-photon drive. This is because restricting the Hilbert space for our model to single excitation subspace, essentially truncates the cavity levels and creates an effective two level system. However, to solve the above equation for the excitation amplitude of the cavity with non-zero $g$, we will differentiate it twice, using the Leibniz integral rule successively to differentiate under the integral sign on the RHS. Doing this allows us to recast Eq.~\eqref{eq:cav_dyn2} as a second order inhomogeneous ODE of the form

\begin{align}
\frac{d^{2}}{dt^{2}}\psi_{\mathrm{C}}\left(t\right)+c_{1}\frac{d}{dt}\psi_{\mathrm{C}}\left(t\right)+c_{0}\psi_{\mathrm{C}}\left(t\right)= & F\left(t\right)\label{eq:Forced_damped_oscillator}
\end{align}

where the coefficients and the inhomogeneous driving term are given
by

\begin{align}
c_{1}= & \frac{\mathcal{K}}{2}-i\Delta_{\mathrm{AC}}\\
c_{0}= & g^{2}-\frac{i\Delta_{\mathrm{AC}}\mathcal{K}}{2}\\
F\left(t\right)= & -\sqrt{\mathcal{K}}\left(\frac{d}{dt}\xi\left(t\right)-i\Delta_{\mathrm{AC}}\xi\left(t\right)\right)
\end{align}

The ODE for the cavity excitation amplitude can be solved by standard methods for inhomogeneous ODEs. The expressions for the cavity and atomic excitation amplitudes may be written as the convolution of the input pulse with some Green's functions as

\begin{align}
\psi_{\mathrm{C}}\left(t\right)= & \int_{-\infty}^{\infty}G_{\mathrm{C}}\left(t-t_{1}\right)\xi\left(t_{1}\right)dt_{1}\\
\psi_{\mathrm{A}}\left(t\right)= & ge^{-i\Delta_{\mathrm{AC}}t}\int_{-\infty}^{\infty}dt_{1}G_{\mathrm{A}}\left(t-t_{1}\right)\xi\left(t_{1}\right)
\end{align}

where the cavity Green's function is given by
\begin{align}
G_{\mathrm{C}}\left(\tau\right)= & G_{\mathrm{C}}^{\left(+\right)}\left(\tau\right)-G_{\mathrm{C}}^{\left(-\right)}\left(\tau\right)
\end{align}

and
\begin{align}
G_{\mathrm{C}}^{\left(\pm\right)}\left(\tau\right)= & \Theta\left(\tau\right)\frac{\sqrt{\mathcal{K}}}{2\Phi}e^{i\Delta_{\mathrm{AC}}\tau}\left(\alpha\pm\Phi\right)e^{-\frac{\left(\alpha\pm\Phi\right)}{2}\tau}
\end{align}

where we have defined the parameters $\alpha=\frac{\mathcal{K}}{2}+i\Delta_{\mathrm{AC}}$ and $\Phi=\sqrt{\alpha^{2}-4g^{2}}$. The atomic Green's function is also found to be of the form

\begin{align}
G_{\mathrm{A}}\left(\tau\right)= & G_{\mathrm{A}}^{\left(+\right)}\left(\tau\right)-G_{\mathrm{A}}^{\left(-\right)}\left(\tau\right)
\end{align}

where the positive and negative atom greens functions can be related to the cavity greens functions as

\begin{align*}
G_{\mathrm{A}}^{\left(\pm\right)}\left(\tau\right)= & \frac{2}{\left(\alpha\pm\Phi\right)}G_{\mathrm{C}}^{\left(\pm\right)}\left(\tau\right).
\end{align*}

We note that the decomposition of the Green's function into +ve and -ve parts appears due to the splitting of the JC system into two polaritons. It can be seen that setting the JC coupling $g$ to zero would kill the -ve Green's function as we would have $\alpha=\Phi$. Although we know the time evolution of the atom and the cavity, the most important element for spectroscopic purposes is the pulse, which will eventually be measured. The time evolution of the pulse shape is found to be

\begin{align}
\tilde{\psi}_{\mathrm{P}}\left(t,\tau\right) & =\int_{-\infty}^{\infty}\left[\delta\left(\tau-\tau_{1}\right)-\sqrt{\mathcal{K}}\Theta\left(t-\tau\right)G_{\mathrm{C}}\left(\tau-\tau_{1}\right)\right]\xi\left(\tau_{1}\right)d\tau_{1}\label{eq:output_pulse_shape}
\end{align}

where $t$ is the actual time variable corresponding to the time evolution and $\tau$ is the temporal index for the temporal shape of the pulse at time $t$. At asymptotic time, for non-zero $\mathcal{K}$ the excitation completely decays out of the JC system and is deposited into the pulse modes. The pulse shape at asymptotic time can then be written in terms of a Green's function for the pulse
\begin{align}
\psi_{\mathrm{P}}^{\infty}\left(\tau\right)= & \int_{-\infty}^{\infty}G_{\mathrm{P}}\left(\tau-\tau_{1}\right)\xi\left(\tau_{1}\right)d\tau_{1}
\end{align}

Note that the tilde has been omitted, since the total density matrix in the asymptotic limit is a pure single-photon state in the pulse degrees of freedom. The state of the asymptotic output pulse is given by

\begin{align}
\left|\psi_{\mathrm{P}}^{\infty}\right\rangle = & \int_{-\infty}^{\infty}d\tau\psi_{\mathrm{P}}^{\infty}\left(\tau\right)a^{\dag}\left(\tau\right)\left|0^{a}\right\rangle 
\end{align}

and the pulse Green's function governing the output pulse shape is given by

\begin{align}
G_{\mathrm{P}}\left(\tau\right)= & \delta\left(\tau\right)-\sqrt{\mathcal{K}}G_{\mathrm{C}}\left(\tau\right)
\end{align}

Now, relabelling the input and output pulse shapes at times $\left(t\rightarrow\pm\infty\right)$ as $\xi_{\mathrm{in/out}}$, we can write from Eq.~\eqref{eq:output_pulse_shape},

\begin{align}
\xi_{\mathrm{out}}\left(\tau\right)= & \xi_{\mathrm{in}}\left(\tau\right)-\sqrt{\mathcal{K}}\int_{-\infty}^{\infty}d\tau G_{\mathrm{C}}\left(\tau-\tau_{1}\right)\xi_{\mathrm{in}}\left(\tau_{1}\right)\label{eq:TD_OP}
\end{align}

The integral is essentially a convolution of the input pulse shape with the cavity Green's function. Moving to the frequency domain via a  transform $\tilde{f}\left(\omega\right)=\int_{-\infty}^{\infty}d\tau f\left(\tau\right)e^{i\omega\tau}/\sqrt{2\pi}$ converts the convolution of two functions in $\tau$-space to a product of two functions in $\omega$-space. Thus under a  transform, the above equation transforms as

\begin{align}
\tilde{\xi}_{\mathrm{out}}\left(\omega\right)= & \left[1-\sqrt{2\pi\mathcal{K}}\tilde{G}_{\mathrm{C}}\left(\omega\right)\right]\tilde{\xi}_{\mathrm{in}}\left(\omega\right)\label{eq:freq_output_shape}
\end{align}

It can be shown that the factor of $1-\sqrt{2\pi\mathcal{K}}\tilde{G}_{\mathrm{C}}\left(\omega\right)$ can be written as a phase factor acting on the input pulse shape. Thus the interaction with the atom-cavity picture only induces a phase onto the spectral amplitude of the input pulse. This implies that the output pulse can be written as a unitary acting on the input pulse
\begin{align}
\left|\xi_{\mathrm{out}}\right\rangle = & U_{\mathrm{P}}\left|\xi_{\mathrm{in}}\right\rangle \label{eq:Freq_domain_output_pulse}
\end{align}
where the pulse unitary is given by expression
\begin{align}
U_{\mathrm{P}} & =\int_{-\infty}^{\infty}d\omega\exp\left\{ -i\varphi\left(\omega\right)\right\} \left|1_{\omega}\right\rangle \left\langle 1_{\omega}\right|\label{eq:Pulse_unitary}
\end{align}

with $\left|1_{\omega}\right\rangle =a^{\dag}\left(\omega\right)\left|0\right\rangle $
and the phase is given by $\varphi=2\tan^{-1}\left[\zeta_{\mathrm{P}}\left(\omega\right)\right]$.
$\zeta_{\mathrm{P}}\left(\omega\right)$ is related to the response of the JC system and is given by

\begin{align}
\zeta_{\mathrm{P}}\left(\omega\right)= & \frac{\mathcal{K}\left(\omega+\Delta_{\mathrm{AC}}\right)}{2\left(\omega^{2}+\omega\Delta_{\mathrm{AC}}-g^{2}\right)}
\end{align}

The output pulse can be written as a unitary acting on the input pulse because there are no losses considered in this model and because the initial excitation is only in the input pulse. Having this unitary structure simplifies the calculations of the FI and allows us to analytically calculate bounds on the maximum precision that can be obtained for our model of cavity-enhanced spectroscopy. We also note that the unitary $U_{\mathrm{P}}$ can be generated by an effective dimensionless Hamiltonian $H_{\mathrm{P}}$, that is given by

\begin{align}
H_{\mathrm{P}}= & 2\int_{-\infty}^{\infty}d\omega\tan^{-1}\left[\zeta_{\mathrm{P}}\left(\omega\right)\right]\left|1_{\omega}\right\rangle \left\langle 1_{\omega}\right|\label{eq:Unitary generator}
\end{align}

\section{QFI of the output pulse\label{QFI_app}}

From Eq.~\eqref{eq:QFI_Gen_out}, we can write down the QFI of the output light at arbitrary time. However, in the infinite time limit, all excitations will dissipate out of the JC system and we can express the state of the output light as a pure single-photon state  $\rho_{\theta,\infty}^{\mathrm{out}}=\left|\xi_{\mathrm{out}}\right\rangle \langle \xi_{\mathrm{out}}|=\left|\tilde{\psi}_{\mathrm{P}}\left(t\rightarrow\infty\right)\right\rangle\left\langle \tilde{\psi}_{\mathrm{P}}\left(t\rightarrow\infty\right) \right| $. In this limit, the first term in Eq.~\eqref{eq:QFI_Gen_out} vanishes as $p_{\theta}$ goes to zero, and we are left with the following expression.

\begin{align}
\mathcal{Q}\left(\theta;\rho_{\theta,\infty}^{\mathrm{out}}\right)= & 4\left[\left\langle \partial_{\theta}\xi_{\mathrm{out}}\left|\partial_{\theta}\xi_{\mathrm{out}}\right.\right\rangle -\left|\left\langle \partial_{\theta}\xi_{\mathrm{out}}\left|\xi_{\mathrm{out}}\right.\right\rangle \right|^{2}\right].\label{eq:Pure_state_QFI}
\end{align}

In the frequency domain, the output light states are given by Eqs.~\eqref{eq:Freq_domain_output_pulse} and~\eqref{eq:Pulse_unitary}. The unitary $U_{\mathrm{P}}$ encodes information about the system parameters into the input pulse. Assuming that the input pulse is independent of these parameters, we have

\begin{align}
\left|\partial_{\theta}\xi_{\mathrm{out}}\right\rangle = & \left(\partial_{\theta}U_{\mathrm{P}}\right)\left|\xi_{\mathrm{in}}\right\rangle .
\end{align}
Now, from the definition of the unitary in Eq.~\eqref{eq:Pulse_unitary}, we have

\begin{align}
\partial_{\theta}U_{\mathrm{P}}=  \int_{-\infty}^{\infty}d\omega\partial_{\theta}\left(\exp\left\{ -2i\tan^{-1}\left[\zeta_{\mathrm{P}}\left(\omega\right)\right]\right\} \right)\left|1_{\omega}\right\rangle \left\langle 1_{\omega}\right|
=  -i\int_{-\infty}^{\infty}d\omega\mathcal{Z}_{\theta}\left(\omega\right)\exp\left\{ -2i\tan^{-1}\left[\zeta_{\mathrm{P}}\left(\omega\right)\right]\right\} \left|1_{\omega}\right\rangle \left\langle 1_{\omega}\right|,
\end{align}
where we have defined the phase sensitivity of the unitary $\mathcal{Z}_{\theta}$ as

\begin{align}
\mathcal{Z}_{\theta}\left(\omega\right)= & \partial_{\theta}\left\{ 2\tan^{-1}\left[\zeta_{\mathrm{P}}\left(\omega\right)\right]\right\} =\frac{2\partial_{\theta}\zeta_{\mathrm{P}}\left(\omega\right)}{1+\left(\zeta_{\mathrm{P}}\left(\omega\right)\right)^{2}}.
\end{align}
The $\theta$-derivative of the output pulse in the frequency domain is then given by
\begin{align}
\left|\partial_{\theta}\xi_{\mathrm{out}}\right\rangle = & -i\int_{-\infty}^{\infty}d\omega\mathcal{Z}_{\theta}\left(\omega\right)\exp\left\{ -i\Phi_{\mathrm{P}}\left(\omega\right)\right\} \tilde{\xi}_{\mathrm{in}}\left(\omega\right)\left|1_{\omega}\right\rangle, 
\end{align}
where we have defined the phase as $\Phi_{\mathrm{P}}\left(\omega\right)=2\tan^{-1}\left[\zeta_{\mathrm{P}}\left(\omega\right)\right]$. The norm of the above ket and its overlap with the output pulse is
then given by

\begin{align}
\left\langle \partial_{\theta}\xi_{\mathrm{out}}\left|\partial_{\theta}\xi_{\mathrm{out}}\right.\right\rangle = & \int_{-\infty}^{\infty}d\omega\left[\mathcal{Z}_{\theta}\left(\omega\right)\right]^{2}\left|\tilde{\xi}_{\mathrm{in}}\left(\omega\right)\right|^{2}\\
\left\langle \partial_{\theta}\xi_{\mathrm{out}}\left|\xi_{\mathrm{out}}\right.\right\rangle = & +i\int_{-\infty}^{\infty}d\omega\mathcal{Z}_{\theta}\left(\omega\right)\left|\tilde{\xi}_{\mathrm{in}}\left(\omega\right)\right|^{2}
\end{align}
Substituting these expressions into Eq.~\eqref{eq:Pure_state_QFI}, we obtain the final expression for the asymptotic QFI for any given input single-photon pulse

\begin{align}
\mathcal{Q}\left(\theta;\rho_{g,\infty}^{\mathrm{out}}\right)= & 4\left[\int_{-\infty}^{\infty}\left[\mathcal{Z}_{\theta}\left(\omega\right)\right]^{2}\left|\tilde{\xi}_{\mathrm{in}}\left(\omega\right)\right|^{2}d\omega-\left|\int_{-\infty}^{\infty}\mathcal{Z}_{\theta}\left(\omega\right)\left|\tilde{\xi}_{\mathrm{in}}\left(\omega\right)\right|^{2}d\omega\right|^{2}\right].\label{eq:QFI_pulse_gen}
\end{align}

We note that the above formula may also be written in terms the $\theta$-derivative of the generator of the pulse unitary written in Eq.~\eqref{eq:Unitary generator}. The derivative of this generator is given by

\begin{align}
\partial_{\theta}H_{\mathrm{P}}=  \partial_{\theta}\int_{-\infty}^{\infty}d\omega2\tan^{-1}\left[\zeta_{\mathrm{P}}\left(\omega\right)\right]\left|1_{\omega}\right\rangle \left\langle 1_{\omega}\right|
=  \int_{-\infty}^{\infty}d\omega\mathcal{Z}_{\theta}\left(\omega\right)\left|1_{\omega}\right\rangle \left\langle 1_{\omega}\right|
\end{align}

The expectation value of $\left(\partial_{\theta}H_{\mathrm{P}}\right)^{2}$ and $\partial_{\theta}H_{\mathrm{P}}$ wrt the input pulse state $\left|\xi_{\mathrm{in}}\right\rangle $, gives us the two terms of the QFI expression, allowing us to write the QFI in a form that is basis-independent

\begin{align}
\mathcal{Q}\left(\theta;\rho_{g,\infty}^{\mathrm{out}}\right)=  4\left[\left\langle \xi_{\mathrm{in}}\right|\left(\partial_{\theta}H_{\mathrm{P}}\right)^{2}\left|\xi_{\mathrm{in}}\right\rangle -\left|\left\langle \xi_{\mathrm{in}}\right|\partial_{\theta}H_{\mathrm{P}}\left|\xi_{\mathrm{in}}\right\rangle \right|^{2}\right]
=  4\mathrm{Var}\left(\partial_{\theta}H_{\mathrm{P}}\right)
\end{align}
and
\begin{align}
\mathcal{Q}_{\mathcal{G}}\left(\left|\psi_{\mathrm{in}}^{\mathrm{P}}\right\rangle \right)=  4\left[\left\langle \psi_{\mathrm{in}}^{\mathrm{P}}\right|\mathcal{Z_{G}}^{2}\left|\psi_{\mathrm{in}}^{\mathrm{P}}\right\rangle -\left|\left\langle \psi_{\mathrm{in}}^{\mathrm{P}}\right|\mathcal{Z_{G}}\left|\psi_{\mathrm{in}}^{\mathrm{P}}\right\rangle \right|^{2}\right]
= 4\mathrm{Var}\left(\mathcal{Z}_{\mathcal{G}}\right)
  \mathcal{Z}_{\mathcal{G}}\left(\omega\right)=\int_{-\infty}^{\infty}d\omega\mathcal{Z}_{\mathcal{G}}\left(\omega\right)\left|1_{\omega}\right\rangle \left\langle 1_{\omega}\right|.
\end{align}
Using the above formulae, the asymptotic QFI can be calculated for the estimation of $\mathcal{G}$ and $\Delta$, for different pulse shapes. However, before we compute the QFI, we will look at the upper bound for QFI that can be obtained using the unitary that generates the scattered pulse. 

\subsection{Bounds on the Asymptotic QFI\label{QFI_bounds deriv}}

The bounds for the QFI at asymptotic time can be calculated from the maximum and minimum eigenvalues of the Hamiltonian in Eq.~\eqref{eq:Unitary generator}  that generates the unitary, which maps the input light pulse to the output pulse at asymptotic time. 

The bounds for the QFI can be written as

\begin{align}
\mathcal{Q}\left(\theta;\left|\xi_{\mathrm{out}}\right\rangle \right)\leq & \left(\mu_{\theta}^{\mathrm{max}}-\mu_{\theta}^{\mathrm{min}}\right)^{2}
\end{align}

where $\mu_{\theta}^{\mathrm{max/min}}$ are the global maxima and minima of the phase sensitivity function $\mathcal{Z}_{\mathcal{G}}\left(\omega\right)$. In principle, the global maxima and minima of the above function can be written down analytically as some function of the system parameters and this would allow us to write the bounds from the expressions for $\mathcal{Z}_{\mathcal{G}}\left(\omega\right)$. However the extremum points for $\mathcal{Z}_{\mathcal{G}}$ are found to be the roots of a quartic  equation with the coefficients having some functional-dependence on the system parameters. Wrt a new coordinate, shifted by the detuning, $\Omega=\omega+\Delta$, the equation for the extremum points of $\mathcal{Z}_{\mathcal{G}}$ is given by

\begin{align}
12\Omega^{4}-16\Delta\Omega^{3}+\left(\mathcal{K}^{2}+4\Delta^{2}-8\mathcal{G}\right)\Omega^{2}-4\mathcal{G}^{2}= & 0\label{eq:Extremum points}
\end{align}

The analytical roots of this quartic equation is quite complicated even for constant coefficients, and for our complicated parameter dependent coefficients, the algebra becomes untractable, which makes the final expressions difficult to understand. It is more instructive to look at these bounds in some physical limits of interest for the parameters in consideration.

We note that Eq.~\eqref{eq:Extremum points} becomes biquadratic in the resonant limit, i.e $\Delta\rightarrow0$, which allows for an analytic solution. In the resonance limit, the shifted frequency coordinate $\Omega$ coincides with the original $\omega$ and we can write the extremum condition as

\begin{align}
12\omega^{4}+\left(\mathcal{K}^{2}-8\mathcal{G}\right)\omega^{2}-4\mathcal{G}^{2}= & 0\label{eq:Res_JC_extremum}
\end{align}

The roots of the above equation are then

\begin{align}
\left(\omega^{2}\right)_{\pm}= & \frac{8\mathcal{G}-\mathcal{K}^{2}\pm\sqrt{\left(\mathcal{K}^{2}-8\mathcal{G}\right)^{2}+192\mathcal{G}^{2}}}{24}
\end{align}

We note that the discriminant here is always positive and larger than $8\mathcal{G}-\mathcal{K}^{2}$. Thus the root $\left(\omega^{2}\right)_{-}$ is negative. Since we require that $\omega^{2}$ must be positive, we take only the +ve roots of $\omega^{2}$. The extremum positions are thus given by $\omega_{\pm}=\pm\sqrt{\omega_{+}^{2}}$. We note that in the resonant limit, the $\mathcal{Z}_{\mathcal{G}}$ function we want to maximise becomes $\mathcal{Z}_{\mathcal{G},\mathrm{res}}\left(\omega\right)=4\mathcal{K}\omega/\left\{ 4\left(\omega^{2}-\mathcal{G}\right)^{2}+\mathcal{K}^{2}\omega^{2}\right\} $. Within this resonant limit, we will now look at two different physical limits for the coupling parameters we have in our system. 

\subsubsection{Bounds on QFI for resonant JC system}

\paragraph{\textbf{Bounds on QFI of $\mathcal{G}$ in the bad cavity limit $\left(\mathcal{\mathcal{K}}^{2}\gg\mathcal{G}\right)$}}

In the bad cavity limit, the rate of coherent exchange between the atom and the cavity is much smaller than the rate at which photons are lost from the cavity. The cavity thus primarily acts as a channel for the excitation and dissipation of the atom via some engineered modes which are coupled to the cavity. We note that our original Hamiltonian assumes that the free space decay rate of the atom is much smaller than engineered interaction rates $g$ and $\kappa$. This ensures that only the decay into these engineered modes, such as some waveguide coupled to the cavity, is significant and this photon leakage effectively suppresses the coherent exchange of energy between the atom and the cavity. Due to these effects, this limit has been studied theoretically within the context of the Purcell effect and experimentally for the purpose of developing single-photon sources, where the goal is to ensure that the photon is quickly emitted into the desired modes. Within the bad cavity limit of the resonant JC model, the extremum points of $\mathcal{Z}_{\mathcal{G},\mathrm{res}}\left(\omega\right)$ are found to be at

\begin{align}
\omega_{\pm}= & \pm2\frac{\mathcal{G}}{\mathcal{K}}
\end{align}

These points are found by a binomial expansion of the discriminant in Eq.~\eqref{eq:Res_JC_extremum} and discarding terms of higher order than $g^{2}/\kappa^{2}$. Thus the maximum and minimum values of $\mathcal{Z}_{\mathcal{G},\mathrm{res}}\left(\omega\right)$ are given by

\begin{align}
\mathcal{Z}_{\mathcal{G},\mathrm{res}}\left(\omega_{\pm}\right)= & \pm\frac{1}{\mathcal{G}}
\end{align}

The bounds on the QFI at aysmptotic time in the very bad cavity limit is thus given by $\mathcal{Q}_{\mathcal{G}}^{\infty}\leq\left(\left.\mathcal{Z}_{\mathcal{G},\mathrm{res}}\right|_{\mathrm{max}}-\left.\mathcal{Z}_{\mathcal{G},\mathrm{res}}\right|_{\mathrm{min}}\right)^{2}=\left(\frac{2}{\mathcal{G}}\right)^{2}$. The bound on the dimensionless asymptotic QFI in this limit is thus found to be 

\begin{align}
\mathcal{G}^{2}\mathcal{\mathcal{Q}_{\mathcal{G}}^{\infty}}\leq & 4
\end{align}

This is the same bound that is found when the atom is in free space. Physically this makes sense as the bad cavity limit is the case when the cavity can be adiabatically eliminated to equate the total system to the dissipation of the atom into some modes with an enhanced decay rate. Therefore the bound on the total information about the atom that can be extracted when it is in a bad cavity is equal to the bound on the information when it is in free space.

\paragraph{\textbf{Bounds on QFI of $\mathcal{G}$ in the JC strong coupling limit $\left(\mathcal{G}\gg\mathcal{\mathcal{K}}^{2}\right)$}}

The JC strong coupling limit is reached when the rate of coherent exchange between the atom and the cavity is much larger than the dissipation rates of the atom and the cavity. It is to be noted that this is strong coupling only within the context of the JC model as the atom-cavity interaction strength is still negligible compared to the transition frequencies of the atom and the cavity. The strong coupling JC limit is the regime where interesting phenomenon such as the formation of polaritons, which are entangled dressed states of the bare atom and cavity states, can be observed experimentally through the vacuum Rabi splitting. Within this strong coupling limit, the extremum points of $\mathcal{Z}_{\mathcal{G},\mathrm{res}}$ are found to be 

\begin{align}
\omega_{\pm}\approx & \pm\sqrt{\mathcal{G}}
\end{align}

Thus, in this limit, the maximum and minimum values of $\mathcal{Z}_{\mathcal{G},\mathrm{res}}\left(\omega\right)$ are given by

\begin{align}
\mathcal{Z}_{\mathcal{G},\mathrm{res}}\left(\pm\sqrt{\mathcal{G}}\right)= & \pm\frac{4}{\mathcal{K}\sqrt{\mathcal{G}}}
\end{align}

The bound on the asymptotic QFI for $\mathcal{G}$-estimation in this limit is then given by $\mathcal{Q}_{\mathcal{G},\mathrm{res}}^{\infty}\leq\left(\left.\mathcal{Z}_{\mathcal{G},\mathrm{res}}\right|_{\mathrm{max}}-\left.\mathcal{Z}_{\mathcal{G},\mathrm{res}}\right|_{\mathrm{min}}\right)^{2}=\left(\frac{8}{\mathcal{K}\sqrt{\mathcal{G}}}\right)^{2}$. Thus, the bound on the asymptotic dimensionless QFI is given by

\begin{align}
\mathcal{G}^{2}\mathcal{Q}_{\mathcal{G},\mathrm{res}}^{\infty}\leq & 64\frac{\mathcal{G}}{\mathcal{K}^{2}}
\end{align}

For non-zero $\mathcal{K}$, we see that the bound on the information that can be extracted from the system at asymptotic time is proportional to the square of the ratio of the atom-cavity coupling $g$ to the cavity-pulse coupling $\mathcal{K}$. This implies the higher the coherent exchange rate is, compared to the rate of photon loss from the cavity, the higher the amount of information that can be extracted about the atom at asymptotic time. For this physical interpretation, it is imperative that $\mathcal{K}$ not be zero, as a non-zero value of $\mathcal{K}$, no matter how small, ensures that a photon enters the cavity at some point and will leave the cavity at asymptotic time. However, at $\mathcal{K}=0$, this physical interpretation does not work, as the initial conditions for which our dynamical equations were solved will give some dynamics only if the cavity-pulse interaction is non-zero. Since the atom and the cavity initially had no excitations, having a zero value of $\mathcal{K}$ means having no dynamics

\subsubsection{Bounds on QFI for detuned JC system}

\begin{figure}[t]
    \centering
    \includegraphics[scale=0.56]{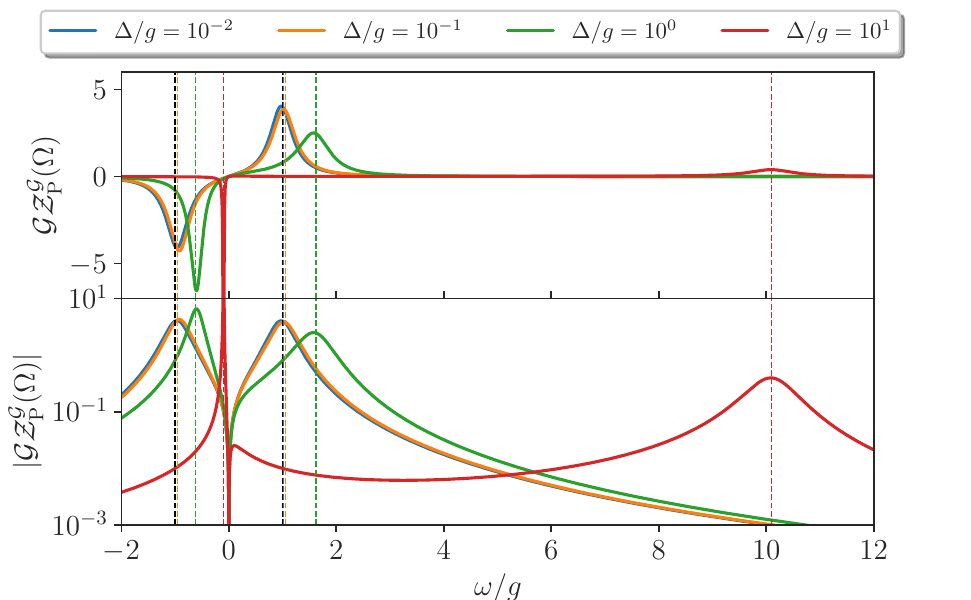}
    \caption{The phase sensitivity function out of resonance plotted for different values of $\Delta/g$. The value of the JC coupling strength was taken to be $g=3.3\times 10^{-4} \mathrm{cm}^{-1}\approx 10 \mathrm{MHz}$ }
    \label{fig:Z_G_P_detuned}
\end{figure}

Although the quartic equation in Eq.~\eqref{eq:Extremum points} cannot be exactly solved to obtain some useful analytical expressions for the general bounds, we can obtain some approximate bounds on the asymptotic QFI in certain limits even for a detuned JC system. To do this, we will look at the nature of the phase sensitivity written in terms of the shifted coordinate $\Omega$

\begin{align}
\mathcal{Z}_{\mathcal{G}}\left(\Omega\right)= & \frac{4\mathcal{K}\Omega}{4\left[A(\Omega)\right]^{2}+\mathcal{K}^{2}\Omega^{2}}=\frac{\mathcal{N}\left(\Omega\right)}{\mathcal{D}\left(\Omega\right)}\label{eq: shifted phase sensi}
\end{align}

where $A\left(\Omega\right)$ is a quadratic polynomial given by $A\left(\Omega\right)=\Omega^{2}-\Omega\Delta-\mathcal{G}$. Plots of the function (Shown in Fig.~\ref{fig:Z_G_P_detuned}) show that there are two extremum points. These extremum points of $\mathcal{Z}_{\mathcal{G}}$ are given by the point where its $\omega$-derivative becomes zero, i.e. $\frac{d\mathcal{Z}_{\mathcal{G}}}{d}=\frac{\mathcal{N}^{\prime}\mathcal{D}-\mathcal{N}\mathcal{D}^{\prime}}{\mathcal{D}^{2}}=0$. Calculating the derivatives of the numerator and the denominator explicitly, the extremum condition in Eq.~\eqref{eq:Extremum points} can be re-written as

\begin{figure}[t]
    \centering
    \includegraphics[scale=0.53]{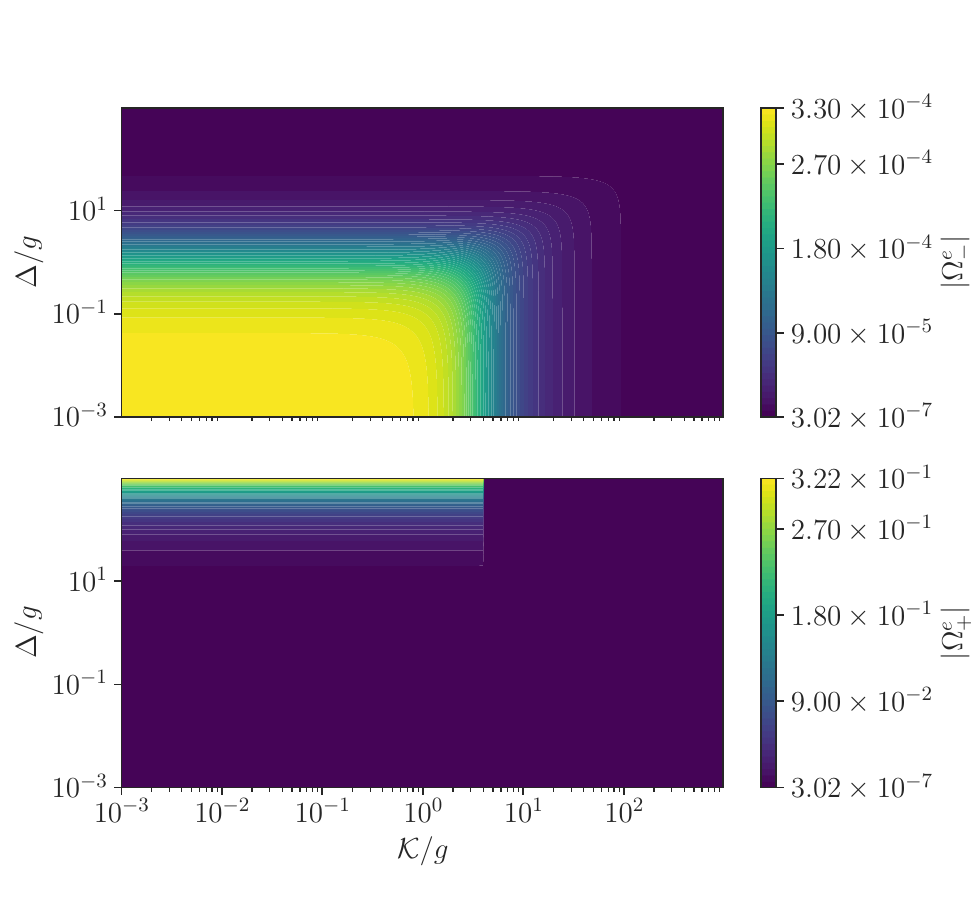} \includegraphics[scale=0.53]{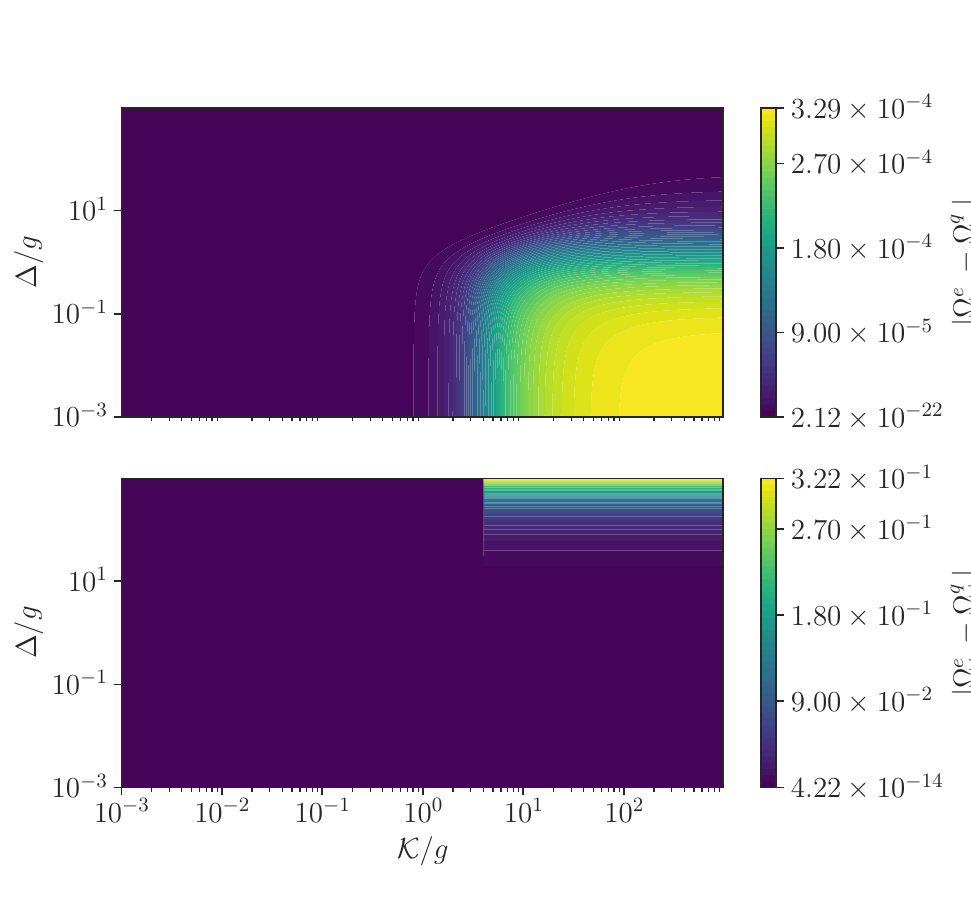} 
    
    \caption{The extremum points of the phase sensitivity function plotted for different values of $\mathcal{K}/g$ and $\Delta/g$ in the left panel. The difference between the exact extremum points and the approximate points $\Omega^q_{\pm}$ is shown in the right panel. The value of the JC coupling strength was taken to be $g=3.3\times 10^{-4} \mathrm{cm}^{-1}\approx 10 \mathrm{MHz}$ }
    \label{fig:Roots}
\end{figure}

\begin{align}
4A\left(\Omega\right)\left\{ 3\Omega^{2}-\Delta\Omega+\mathcal{G}\right\} +\mathcal{K}^{2}\Omega^{2}= & 0\label{eq:Extremum cond}
\end{align}

This is still a quartic equation, whose exact solutions $\Omega=\Omega_{\pm}^{e}$ can be obtained numerically (Shown in Fig.~\ref{fig:Roots}). However writing them down analytically in a closed form is difficult due to the quartic nature. To get around this, we will consider some approximate solutions in two different limits of the JC model, just as we did for the resonant case. 

\paragraph{\textbf{Strongly Coupled JC system $\left(\mathcal{K}\ll\Omega_{\mathrm{R}}\right)$}}

We note that for values of $\mathcal{K}$ that are much smaller than the other coupling parameters, the quadratic term $A\left(\Omega\right)$ will dominate the values of the denominator. Thus in this regime, the stationary points $\Omega_{\pm}^{e}$ which are the roots of Eq.~\eqref{eq:Extremum cond} will be near the roots of the quadratic term $A\left(\Omega\right)$, which are found to be 

\begin{align}
\Omega_{\pm}^{q}= & \frac{\Delta\pm\Omega_{\mathrm{R}}}{2}
\end{align}

where $\Omega_{\mathrm{R}}=\sqrt{\Delta^{2}+4\mathcal{G}}$ is the Rabi frequency of the detuned JC system. The difference between the actual extremum points and the roots of the quadratic equation are plotted in Fig.~\ref{fig:Roots} and we see that the difference is very small in the region where $\mathcal{K}$ is small. Thus in this region, we can write the exact extremum points as

\begin{align}
\Omega_{\pm}^{e}\approx & \Omega_{\pm}^{q}
\end{align}

This allows us to write an analytical expression for the bounds for the detuned JC system under strong coupling. The maximum and minimum values of the phase sensitivity function can be analytically written as

\begin{align}
\mathcal{Z}_{\mathcal{G}}\left(\Omega_{\pm}^{q}\right)= & \frac{4}{\mathcal{K}\Omega^q_{\pm}}\nonumber \\
= & \frac{8}{\mathcal{K}\left(\Delta\pm\Omega_{\mathrm{R}}\right)}
\end{align}
Therefore, the QFI bounds are obtained as
\begin{align}
\mathcal{Q}_{\mathcal{G}}^{\infty}\leq\mathcal{Q}_{\mathcal{G},\mathrm{max}}^{\infty}\approx & \left(\mathcal{Z}_{\mathcal{G}}\left(\Omega_{+}^{q}\right)-\mathcal{Z}_{\mathcal{G}}\left(\Omega_{-}^{q}\right)\right)^{2}\nonumber \\
= & \left(\frac{8}{\mathcal{K}\left(\Delta+\Omega_{\mathrm{R}}\right)}-\frac{8}{\mathcal{K}\left(\Delta-\Omega_{\mathrm{R}}\right)}\right)^{2}\nonumber \\
= & \left(\frac{8\left\{ \left(\Delta-\Omega_{\mathrm{R}}\right)-\left(\Delta+\Omega_{\mathrm{R}}\right)\right\} }{\mathcal{K}\left(\Delta+\Omega_{\mathrm{R}}\right)\left(\Delta-\Omega_{\mathrm{R}}\right)}\right)^{2}\nonumber \\
= & \frac{64}{\mathcal{K}^{2}}\left(\frac{-2\Omega_{\mathrm{R}}}{\Delta^{2}-\Omega_{\mathrm{R}}^{2}}\right)^{2}=\frac{64}{\mathcal{K}^{2}}\left(\frac{-2\Omega_{\mathrm{R}}}{-4\mathcal{G}}\right)^{2}
\end{align}

\begin{figure}[t]
    \centering
    \includegraphics[scale=0.53]{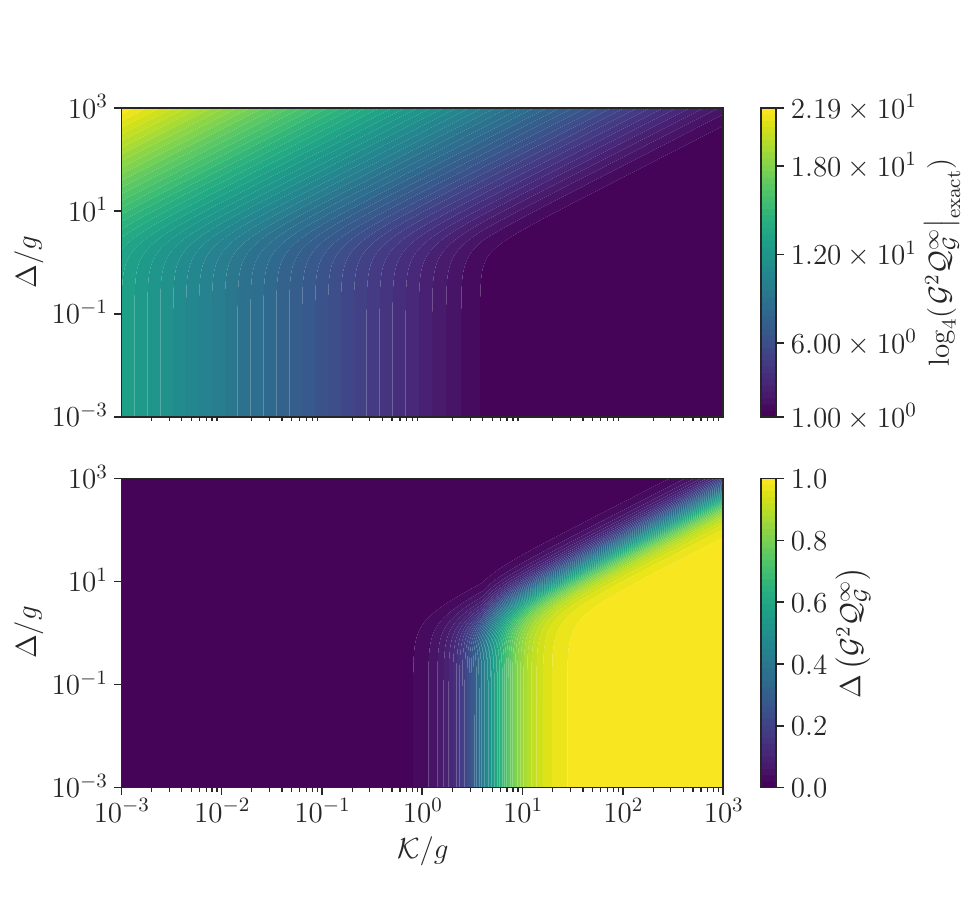} \includegraphics[scale=0.53]{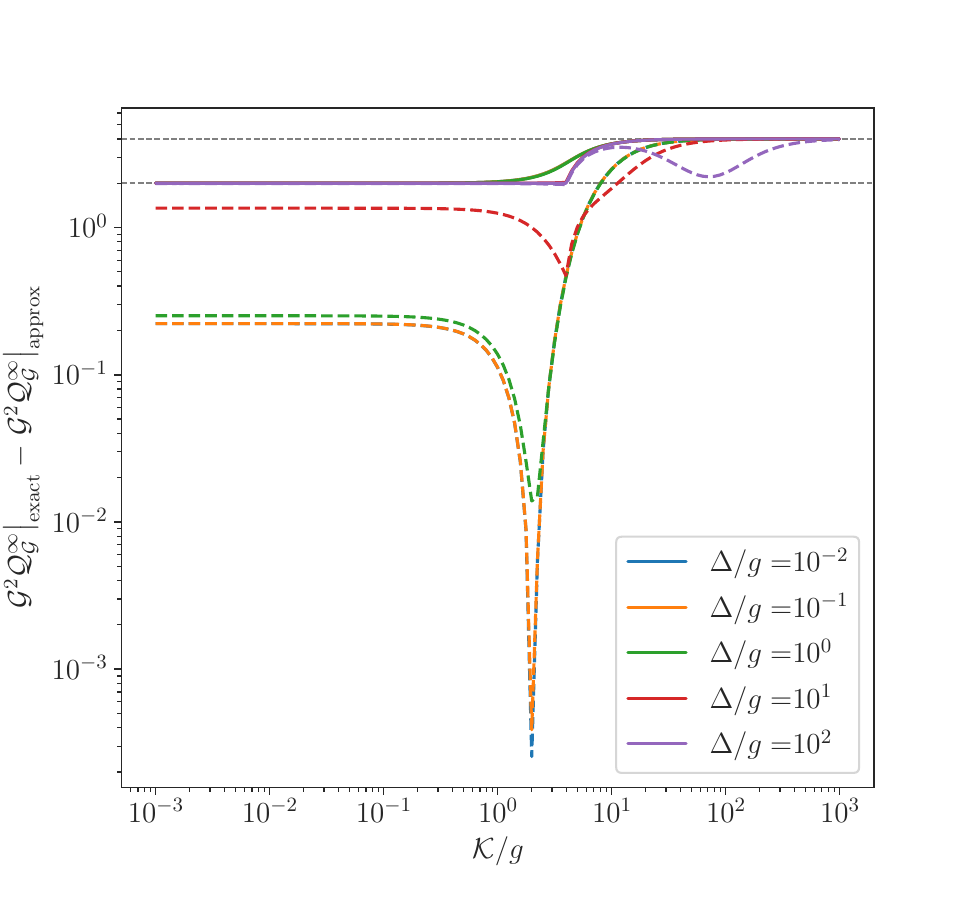} 
    
    \caption{The top left panel shows the exact asymptotic QFI bounds plotted on a $\log_4$ scale for different values of $\mathcal{K}/g$ and $\Delta/g$. The bottom left panel shows the relative error in the approximate asymptotic QFI bounds obtained, for different values of $\mathcal{K}/g$ and $\Delta/g$. The right panel shows the absolute difference between the exact QFI bounds and the approximate QFI bounds plotted against $\mathcal{K}/g$ for a few different values of $\Delta/g$. The solid lines represent the errors for the bounds obtained with the extremum points taken to be at $\Omega^q_\pm$ and the dashed lines represent the errors with the first order corrections taken into account. The value of the JC coupling strength was taken to be $g=3.3\times 10^{-4} \mathrm{cm}^{-1}\approx 10 \mathrm{MHz}$ }
    \label{fig:Bounds_and_errors}
\end{figure}

The bound on the dimensionless QFI in this region of the parameter space is then given by

\begin{align}
\mathcal{G}^{2}\mathcal{Q}_{\mathcal{G}}^{\infty}\leq & 16\frac{\Omega_{\mathrm{R}}^{2}}{\mathcal{K}^{2}}
\end{align}

This expression for the QFI bound allows us to draw the same physical interpretation that we had for the resonant case. The Rabi frequency $\Omega_{\mathrm{R}}$ represents the rate of coherent exchange between the TLA and the cavity mode. The larger this rate of exchange compared to the rate of photon loss $\mathcal{K}$, the larger the maximum attainable QFI becomes. The QFI bounds computed with the numerically obtained exact extremum points $\Omega_{\pm}^{e}$ are shown in Fig.~\ref{fig:Bounds_and_errors}. In the resonant limit, we have $\Omega_{\mathrm{R}}^{2}=4\mathcal{G}$ and the above approximate bound converges to the exact bound we found for the resonant condition, ie $\mathcal{G}^{2}\mathcal{Q}_{\mathcal{G},\mathrm{res}}^{\infty}\leq64\frac{\mathcal{G}}{\mathcal{K}^{2}}$. The relative error in the approximate bounds written above, ie $\Delta\left(\mathcal{G}^{2}\mathcal{Q}_{\mathcal{G}}^{\infty}\right)=\frac{\left\{ \left.\mathcal{G}^{2}\mathcal{Q}_{\mathcal{G}}^{\infty}\right|_{\mathrm{exact}}-\left.\mathcal{G}^{2}\mathcal{Q}_{\mathcal{G}}^{\infty}\right|_{\mathrm{approx}}\right\} }{\left.\mathcal{G}^{2}\mathcal{Q}_{\mathcal{G}}^{\infty}\right|_{\mathrm{exact}}}$ is shown in Fig.~\ref{fig:Bounds_and_errors}. It is observed that the relative error is extremely small in the strong coupling regime, where we are making this approximation.

Although this bound we have found converges to the exact bound in the resonant limit perfectly, we must keep in mind that it is an approximation. Due to this, even though the relative error is negligibly small, we observe a finite absolute error in the approximate QFI bounds obtained, shown in Fig.~\ref{fig:Bounds_and_errors}. The reason for this is that the real extremum points do not exactly coincide with the quadratic roots. Since they lie close to each other in the strong coupling regime, we can assume that the exact roots can be written as some small shift $\delta_{\pm}^{s}$ to the $\Omega_{\pm}^{q}$

\begin{align}
\Omega_{\pm}^{e}= & \Omega_{\pm}^{q}+\delta_{\pm}^{s}
\end{align}

such that 

\begin{align}
\left|\delta_{\pm}^{s}\right|\ll & 1
\end{align}

Under this assumption, we expand the LHS of Eq.~\eqref{eq:Extremum cond} using the above expression for $\Omega_{\pm}^{e}$ and neglect the terms of order higher than $\mathcal{O}\left\{ \left(\delta_{\pm}^{s}\right)^{2}\right\} $. This allows us to obtain the first order correction to the expression for the extremum points as

\begin{align}
\delta_{\pm}^{s}= & \frac{\mathcal{K}^{2}}{\left\{ \pm4\Omega_{\mathrm{R}}\left[\frac{\Delta}{\Omega_{\pm}^{q}}-\frac{\mathcal{G}}{\left(\Omega_{\pm}^{q}\right)^{2}}-3\right]-2\frac{\mathcal{K}^{2}}{\Omega_{\pm}^{q}}\right\} }
\end{align}

Including this first order correction in the calculation of the bound reduces the absolute error in the approximate QFI bound obtained (Shown in Fig.~\ref{fig:Bounds_and_errors}). Thus we observe that including these corrections make our approximation even better. In the strong coupling regime, where $\mathcal{K}$ is small compared to the other parameters, these corrections themselves are quite small, making our approximation quite a good one.

\paragraph{\textbf{Bad Cavity Limit $\left(\mathcal{K}\gg\Omega_{\mathrm{R}}\right)$}}

In the limit, where $\mathcal{K}$ is the very large compared to other parameters, we observe that the extremum points occur around $\Omega=0$  (Shown in Fig.~\ref{fig:Roots}). However they are not exactly at zero, since the phase sensitivity becomes zero at this point. To find where the roots occur around the vicinity of zero, we will assume that their position takes the following form 

\begin{align}
\Omega_{\pm}^{0}= & \frac{x}{\mathcal{K}}
\end{align}

where $x$ has the dimension of $\mathcal{G}$ and is of $\mathcal{O}\left(1\right)$ to ensure that $\Omega_{\pm}^{0}$ are near zero when $\mathcal{K}$ is large. For these values of $\Omega$, the quadratic term becomes 

$A\left(\Omega\right)=-\Delta\frac{x}{\mathcal{K}}-\mathcal{G}+\mathcal{O}\left(\mathcal{K}^{-2}\right)$. The exact condition for the extremum points in Eq.~\eqref{eq:Extremum cond} is then modified as

\begin{align}
4A\left(\Omega\right)\left\{ \mathcal{O}\left(\mathcal{K}^{-2}\right)-\Delta\frac{x}{\mathcal{K}}+\mathcal{G}\right\} +\mathcal{K}^{2}\frac{x^{2}}{\mathcal{K}^{2}}= & 0\nonumber \\
\implies4\left\{ \left(\mathcal{O}\left(\mathcal{K}^{-2}\right)-\Delta\frac{x}{\mathcal{K}}\right)-\mathcal{G}\right\} \left\{ \left(\mathcal{O}\left(\mathcal{K}^{-2}\right)-\Delta\frac{x}{\mathcal{K}}\right)+\mathcal{G}\right\} +x^{2}= & 0\nonumber \\
\implies4\left\{ \left(\mathcal{O}\left(\mathcal{K}^{-2}\right)-\mathcal{O}\left(\mathcal{K}^{-1}\right)\right)^{2}-\mathcal{G}^{2}\right\} +x^{2}= & 0
\end{align}

Discarding terms of the order $\mathcal{O}\left(\mathcal{K}^{-2}\right)$ and smaller, we get

\begin{align}
x^{2}= & 4\mathcal{G}^{2}
\end{align}

Thus the approximate roots near zero in the bad cavity limit are found to be

\begin{align}
\Omega_{\pm}^{0} & =\pm2\frac{\mathcal{G}}{\mathcal{K}}
\end{align}

This is equal to the exact roots we found for the bad cavity limit in the case of a resonant JC system. Substituting these values into the phase sensitivity function and retaining only the terms up to the leading order and neglecting terms of $\mathcal{O}\left(\mathcal{K}^{-1}\right)$, we find

\begin{align}
\mathcal{Z}_{\mathcal{G}}\left(\Omega_{\pm}^{0}\right)= & =\frac{\mathcal{N}\left(\Omega_{\pm}^{0}\right)}{\mathcal{D}\left(\Omega_{\pm}^{0}\right)}=\frac{4\mathcal{K}\Omega}{4\left[A(\Omega)\right]^{2}+\mathcal{K}^{2}\Omega^{2}}
\end{align}

where

\begin{align}
\mathcal{N}\left(\Omega_{\pm}^{0}\right)\approx & 4\mathcal{K}\left(\pm2\frac{\mathcal{G}}{\mathcal{K}}\right)=\pm8\mathcal{G}\\
\mathcal{D}\left(\Omega_{\pm}^{0}\right)\approx & 4\mathcal{G}^{2}+\mathcal{K}^{2}\left(\pm2\frac{\mathcal{G}}{\mathcal{K}}\right)^{2}=8\mathcal{G}^{2}
\end{align}

Therefore, we have

\begin{align}
\mathcal{Z}_{\mathcal{G}}\left(\Omega_{\pm}^{0}\right)\approx & \pm\frac{1}{\mathcal{G}}
\end{align}

Thus, the dimensionless QFI bounds in this regime are found to be given by

\begin{align}
\mathcal{G}^{2}\mathcal{\mathcal{Q}_{\mathcal{G}}^{\infty}} & \leq4
\end{align}

Thus the bounds for the detuned JC system are quite similar to the ones we obtain for the resonant case. Therefore the physical interpretations in the two different limits considered, still hold despite the presence of detuning in the JC system.  We note that if we are to retain terms of $\mathcal{O}\left(\mathcal{K}^{-1}\right)$, the extremum values are found to be $\mathcal{Z}_{\mathcal{G}}\left(\Omega_{\pm}^{0}\right)\approx\pm\frac{1}{\mathcal{G}}\left(1\mp\frac{\Delta}{\mathcal{K}}\right)=\pm\frac{1}{\mathcal{G}}-\frac{\Delta}{\mathcal{K}\mathcal{G}}$. This however, has no effect on the QFI bound as it depends only on the difference between the extremum values and the additional $-\frac{\Delta}{\mathcal{K}\mathcal{G}}$ terms cancel out.

\subsection{Bounds on QFI near polariton frequencies\label{polariton_QFI_bounds}}

The expression for the QFI is given by

\begin{align}
\mathcal{Q}\left(\theta;\rho_{g,\infty}^{\mathrm{out}}\right)= & 4\left[\int_{-\infty}^{\infty}\left[\mathcal{Z}_{\theta}\left(\omega\right)\right]^{2}\left|\tilde{\xi}_{\mathrm{in}}\left(\omega\right)\right|^{2}d\omega-\left|\int_{-\infty}^{\infty}\mathcal{Z}_{\theta}\left(\omega\right)\left|\tilde{\xi}_{\mathrm{in}}\left(\omega\right)\right|^{2}d\omega\right|^{2}\right]\label{eq:QFI_pulse_gen_APP}
\end{align}

\begin{align}
\mathcal{Z}_{\mathcal{G}}\left(\omega\right)= & \frac{4\mathcal{K}\left(\omega+\Delta\right)}{4\left(\omega^{2}+\omega\Delta-\mathcal{G}\right)^{2}+\mathcal{K}^{2}\left(\omega+\Delta\right)^{2}}
\end{align}

Defining $A\left(\omega\right)=\omega^{2}+\omega\Delta-\mathcal{G}$, we know that for small $\mathcal{K}$, the maxima of $\mathcal{Z}_{\mathcal{G}}$ occur at the roots of $A\left(\omega\right)\rightarrow\omega_{\pm}=\frac{1}{2}\left(-\Delta\pm\Omega_{\mathrm{R}}\right)$. For large detunings, it is observed that the peak at $\omega_{-}$ becomes much sharper and taller than the peak at $\omega_{+}$ and the position of these peaks move farther away. The DAE is most pronounced when the input pulse is chosen to be resonant with the maxima at $\omega_{-}$. When the pulse is centred at $\omega_{-}$, we assume that the peak value of the pulse is $P_{0}=\left|\tilde{\xi}_{\mathrm{in}}\left(\omega_{-}\right)\right|^{2}$ and that the pulse is slowly varying compared to the narrow peak at $\omega_{-}$. We also assume that $\left|\tilde{\xi}_{\mathrm{in}}\left(\omega\right)\right|^{2}$ decays sufficiently fast such that the distant peaks at $\omega_{+}$ have negligible contribution to the QFI integrals.

We define $\epsilon\left(\omega\right)=\frac{\mathcal{K}}{2}\left|\omega+\Delta\right|$. In terms of this, the phase sensitivity function can be re-written as

\begin{align}
\mathcal{Z}_{\mathcal{G}}\left(\omega\right)= & 2\text{sgn}\left(\omega+\Delta\right)\frac{\epsilon\left(\omega\right)}{A\left(\omega\right)^{2}+\epsilon\left(\omega\right)^{2}}
\end{align}

In the strong coupling limit, it is reasonable to assume take the limit $\epsilon\rightarrow0$, as $\mathcal{K}$ can be assumed to be very small compared to the other parameters of the system. This allows us to use the Dirac delta distribution limit $\lim_{\epsilon\rightarrow0}\frac{\epsilon}{x^{2}+\epsilon^{2}}=\pi\delta\left(x\right)$. Therefore, in the strong coupling limit, we can write

\begin{align}
\mathcal{Z}_{\mathcal{G}}\left(\omega\right)\approx & 2\text{sgn}\left(\omega+\Delta\right)\lim_{\epsilon\rightarrow0}\frac{\epsilon\left(\omega\right)}{A\left(\omega\right)^{2}+\epsilon\left(\omega\right)^{2}}\\
= & 2\pi\text{sgn}\left(\omega+\Delta\right)\delta\left(A\left(\omega\right)\right)
\end{align}

Now, using the composition rule for the Dirac delta, i.e 
\begin{align}
\delta\left(A\left(\omega\right)\right)= & \sum_{k\in\left\{ +,-\right\} }\frac{\delta\left(\omega-\omega_{k}\right)}{\left|A^{'}\left(\omega_{k}\right)\right|}
\end{align}

It can be shown that $\left|A^{'}\left(\omega_{k}\right)\right|=\Omega_{\mathrm{R}}$.
Thus, the phase sensitivity function can be approximated as

\begin{align}
\mathcal{Z}_{\mathcal{G}}\left(\omega\right)\approx & 2\pi\text{sgn}\left(\omega+\Delta\right)\left[\frac{\delta\left(\omega-\omega_{-}\right)}{\Omega_{\mathrm{R}}}+\frac{\delta\left(\omega-\omega_{+}\right)}{\Omega_{\mathrm{R}}}\right]\label{eq:Z_G_delta_approx}
\end{align}

Now, when we take the expectation value of $\mathcal{Z}_{\mathcal{G}}\left(\omega\right)$ with respect to a pulse that is centred at $\omega_{-}$, only the delta function at $\omega_{-}$ will survive. This is equivalent to assuming that the pulse does not appreciably change within the narrow peak of $\mathcal{Z}_{\mathcal{G}}\left(\omega\right)$ at $\omega_{-}$ and it is slowly varying around this area, but it rapidly decays before it reaches the other peak at $\omega_{+}$. Thus, the expectation value of $\mathcal{Z}_{\mathcal{G}}\left(\omega\right)$ with respect to a pulse that is centred at $\omega_{-}$ is given by

\begin{align}
\left\langle \mathcal{Z}_{\mathcal{G}}\right\rangle _{\omega_{-}}\approx & P_{0}\int_{\omega_{-}}\mathcal{Z}_{\mathcal{G}}\left(\omega\right)d\omega\\
= & -\frac{2\pi P_{0}}{\Omega_{\mathrm{R}}}
\end{align}

The negative sign comes from the sign function, as $\omega_{-}+\Delta=\left(\Delta-\Omega_{\mathrm{R}}\right)/2$, which is always negative. The square of the mean can be written as

\begin{align}
\left\langle \mathcal{Z}_{\mathcal{G}}\right\rangle _{\omega_{-}}^{2}= & \frac{4\pi^{2}P_{0}^{2}}{\Delta^{2}+4\mathcal{G}}
\end{align}

Thus the square of the mean eventually goes to zero as the detuning is increased. Now we look at the square of the phase sensitivity function, which maybe written as

\begin{align}
\mathcal{Z}_{\mathcal{G}}\left(\omega\right)^{2}= & \frac{4\epsilon\left(\omega\right)^{2}}{\left(A\left(\omega\right)^{2}+\epsilon\left(\omega\right)^{2}\right)^{2}}
\end{align}

To approximate this, we will use the un-normalised cauchy representation of the delta function $\lim_{\epsilon\rightarrow0}\frac{\epsilon^{2}}{\left(x^{2}+\epsilon^{2}\right)^{2}}=\frac{\pi}{2\epsilon}\delta\left(x\right)$. Thus in the strong coupling limit, since we can go to the limit $\epsilon\rightarrow0$, the square of the phase sensitivity function can be approximated as

\begin{align}
\mathcal{Z}_{\mathcal{G}}\left(\omega\right)^{2}\approx & 4\frac{\pi}{2\epsilon\left(\omega\right)}\delta\left(A\left(\omega\right)\right)\\
= & \frac{2\pi}{\epsilon\left(\omega\right)}\left[\frac{\delta\left(\omega-\omega_{-}\right)}{\Omega_{\mathrm{R}}}+\frac{\delta\left(\omega-\omega_{+}\right)}{\Omega_{\mathrm{R}}}\right]
\end{align}

Thus, the expectation value of this square of the delta function wrt a pulse centred at $\omega_{-}$ can be approximated as

\begin{align}
\left\langle \left(\mathcal{Z}_{\mathcal{G}}\right)^{2}\right\rangle _{\omega_{-}}\approx & P_{0}\int_{\omega_{-}}\mathcal{Z}_{\mathcal{G}}\left(\omega\right)^{2}d\omega\\
= & \frac{2\pi P_{0}}{\epsilon\left(\omega_{-}\right)\Omega_{\mathrm{R}}}\\
= & \frac{2\pi P_{0}}{\frac{\mathcal{K}}{2}\left|\omega_{-}+\Delta\right|\Omega_{\mathrm{R}}}
\end{align}

Since $\omega_{-}+\Delta=\left(\Delta-\Omega_{\mathrm{R}}\right)/2$, the modulus is given by $\left(\Omega_{\mathrm{R}}-\Delta\right)/2$, as the Rabi frequency is given by $\Omega_{\mathrm{R}}=\sqrt{\Delta^{2}+4\mathcal{G}}$. Thus we have

\begin{align}
\left\langle \left(\mathcal{Z}_{\mathcal{G}}\right)^{2}\right\rangle _{\omega_{-}}\approx & \frac{8\pi P_{0}}{\mathcal{K}\left(\Omega_{\mathrm{R}}-\Delta\right)\Omega_{\mathrm{R}}}\\
= & \frac{8\pi P_{0}\left(\Omega_{\mathrm{R}}+\Delta\right)}{\mathcal{K}\Omega_{\mathrm{R}}\left(\Omega_{\mathrm{R}}^{2}-\Delta^{2}\right)}\\
= & \frac{2\pi P_{0}\left(\Omega_{\mathrm{R}}+\Delta\right)}{\mathcal{K}\Omega_{\mathrm{R}}\mathcal{G}}\\
= & \frac{2\pi P_{0}}{\mathcal{K}\mathcal{G}}\left(1+\frac{\Delta}{\Omega_{\mathrm{R}}}\right)
\end{align}

Thus the expectation value of the square of the phase sensitivity function monotonically increases with increase in detuning and eventually will saturate at very large detunings as $\Delta\gg g\implies\frac{\Delta}{\Omega_{\mathrm{R}}}\approx1$. The rate of increase can be calculated as

\begin{align}
\frac{\partial}{\partial\Delta}\left\langle \left(\mathcal{Z}_{\mathcal{G}}\right)^{2}\right\rangle _{\omega_{-}}= & \frac{2\pi P_{0}}{\mathcal{K}\mathcal{G}}\partial_{\Delta}\left(\frac{\Delta}{\Omega_{\mathrm{R}}}\right)\\
= & \frac{2\pi P_{0}}{\mathcal{K}\mathcal{G}\Omega_{\mathrm{R}}}\left(1-\frac{\Delta^{2}}{\Omega^2_{\mathrm{R}}}\right)
\end{align}

Thus, $\left\langle \left(\mathcal{Z}_{\mathcal{G}}\right)^{2}\right\rangle _{\omega_{-}}$ monotonically increases with increase in detuning as $\Delta/\Omega_{\mathrm{R}}<1$. At asymptotically high detuning however, the ratio $\Delta/\Omega_{\mathrm{R}}$ approaches one and the rate of change goes to zero.  Now, the largest value of variance that can be achieved for the phase sensitivity with a pulse centred at $\omega_{-}$ is therefore set by the largest value of $\left\langle \left(\mathcal{Z}_{\mathcal{G}}\right)^{2}\right\rangle _{\omega_{-}}$ as the mean goes to zero in the large detuning limit. Thus we can write

\begin{align}
\max\left(\mathrm{Var}\left[\mathcal{Z}_{\mathcal{G}}\right]\right)_{\omega_{-}}= & \frac{4\pi P_{0}}{\mathcal{K}\mathcal{G}}
\end{align}

Without taking the asymptotically large detuning limit, the variance is given by

\begin{align}
\mathrm{Var}\left[\mathcal{Z}_{\mathcal{G}}\right]_{\omega_{-}}= & \frac{2\pi P_{0}}{\mathcal{K}\mathcal{G}}\left(1+\frac{\Delta}{\Omega_{\mathrm{R}}}\right)-\left(\frac{2\pi P_{0}}{\Omega_{\mathrm{R}}}\right)^{2}\nonumber \\
= & \frac{2\pi P_{0}}{\mathcal{K}\mathcal{G}}\left(1+\frac{\Delta}{\Omega_{\mathrm{R}}}\right)-\frac{4\pi^{2}P_{0}^{2}}{\Omega_{\mathrm{R}}^{2}}\nonumber \\
= & 4\pi P_{0}\left[\frac{1}{2\mathcal{K}\mathcal{G}}\left(1+\frac{\Delta}{\Omega_{\mathrm{R}}}\right)-\frac{\pi P_{0}}{\Omega_{\mathrm{R}}^{2}}\right]
\end{align}

Thus the maximum QFI that can be obtained via detuning assisted enhancement with a pulse centred at $\omega_{-}$ is capped at

\begin{align}
\mathcal{Q}_{\mathcal{G}}^{\infty}\leq & \frac{16\pi P_{0}}{\mathcal{K}\mathcal{G}}
\end{align}

If the pulse is taken with a peak at $\omega_{+}$ with $P_0=\left|\tilde{\xi}_{\mathrm{in}}\left(\omega_{+}\right)\right|^{2}$, we then have 
\begin{align}
\left\langle \mathcal{Z}_{\mathcal{G}}\right\rangle _{\omega_{+}}\approx & P_{0}\int_{\omega_{+}}\mathcal{Z}_{\mathcal{G}}\left(\omega\right)d\omega\nonumber \\
= & +\frac{2\pi P_{0}}{\Omega_{\mathrm{R}}}
\end{align}

\begin{align}
\left\langle \left(\mathcal{Z}_{\mathcal{G}}\right)^{2}\right\rangle _{\omega_{+}}\approx & P_{0}\int_{\omega_{+}}\mathcal{Z}_{\mathcal{G}}\left(\omega\right)^{2}d\omega\nonumber \\
= & \frac{2\pi P_{0}}{\epsilon\left(\omega_{+}\right)\Omega_{\mathrm{R}}}\nonumber \\
= & \frac{2\pi P_{0}}{\frac{\mathcal{K}}{2}\left|\omega_{+}+\Delta\right|\Omega_{\mathrm{R}}}
\end{align}

We know $\omega_{\pm}=\frac{1}{2}\left(-\Delta\pm\Omega_{\mathrm{R}}\right)\implies\left|\omega_{+}+\Delta\right|=\left|\frac{1}{2}\left(-\Delta+\Omega_{\mathrm{R}}\right)+\Delta\right|=\frac{\Omega_{\mathrm{R}}+\Delta}{2}$. Thus, we have

\begin{align}
\left\langle \left(\mathcal{Z}_{\mathcal{G}}\right)^{2}\right\rangle _{\omega_{+}}\approx & \frac{8\pi P_{0}}{\mathcal{K}\Omega_{\mathrm{R}}\left(\Omega_{\mathrm{R}}+\Delta\right)}\nonumber \\
= & \frac{8\pi P_{0}\left(\Omega_{\mathrm{R}}-\Delta\right)}{\mathcal{K}\Omega_{\mathrm{R}}\left(\Omega_{\mathrm{R}}^{2}-\Delta^{2}\right)}=\frac{8\pi P_{0}\left(\Omega_{\mathrm{R}}-\Delta\right)}{\mathcal{K}\Omega_{\mathrm{R}}\left(4\mathcal{G}\right)}\nonumber \\
= & \frac{2\pi P_{0}\left(\Omega_{\mathrm{R}}-\Delta\right)}{\mathcal{K}\Omega_{\mathrm{R}}\mathcal{G}}=\frac{4\pi P_{0}}{2\mathcal{K}\mathcal{G}}\left(1-\frac{\Delta}{\Omega_{\mathrm{R}}}\right)
\end{align}

The variance is then 

\begin{align}
\left\langle \left(\mathcal{Z}_{\mathcal{G}}\right)^{2}\right\rangle _{\omega_{+}}-\left|\left\langle \mathcal{Z}_{\mathcal{G}}\right\rangle _{\omega_{+}}\right|^{2}\approx & \frac{4\pi P_{0}}{2\mathcal{K}\mathcal{G}}\left(1-\frac{\Delta}{\Omega_{\mathrm{R}}}\right)-\left(\frac{4\pi^{2}P_{0}^{2}}{\Omega_{\mathrm{R}}^{2}}\right)\\
= & 4\pi P_{0}\left(\frac{1}{2\mathcal{K}\mathcal{G}}\left(1-\frac{\Delta}{\Omega_{\mathrm{R}}}\right)-\frac{\pi P_{0}}{\Omega_{\mathrm{R}}^{2}}\right)
\end{align}

For a fixed value of $\mathcal{G}$, the Rabi frequency is minimum at atom-cavity detuning. Thus the variance is maximised when atom cavity detuning is zero. For large detunings, the QFI will vanish for Pulses centred at $\omega_{+}$.

\begin{align}
\mathcal{G}^{2}\mathcal{Q}_{\mathcal{G}}^{+}\approx & 16\pi P_{0}\mathcal{G}^{2}\left[\frac{1}{2\mathcal{K}\mathcal{G}}\left(1-\frac{\Delta}{\Omega_{\mathrm{R}}}\right)-\frac{\pi P_{0}}{\Omega_{\mathrm{R}}^{2}}\right]\\
\mathcal{G}^{2}\mathcal{Q}_{\mathcal{G}}^{-}\approx & 16\pi P_{0}\mathcal{G}^{2}\left[\frac{1}{2\mathcal{K}\mathcal{G}}\left(1+\frac{\Delta}{\Omega_{\mathrm{R}}}\right)-\frac{\pi P_{0}}{\Omega_{\mathrm{R}}^{2}}\right]
\end{align}

It can be seen that, in the resonant limit, both of these converge to $\mathcal{G}^{2}\mathcal{Q}_{\mathcal{G}}^{+}=\mathcal{G}^{2}\mathcal{Q}_{\mathcal{G}}^{-}\approx4\pi P_{0}\mathcal{G}\left(\frac{2}{\mathcal{K}}-\pi P_{0}\right)$. For the normalised pulses we consider, it can be shown  that $P_0=c_{\mathrm{P}}T$, where $T$ is the pulse duration and $c_{\mathrm{P}} $ is a geometric factor that is governed by the pulse shape. For the three pulse shapes we consider, its values are given by
\begin{align}
c_{\mathrm{P}}= & \begin{cases}
\sqrt{\frac{2}{\pi}}\approx0.80, & \text{for Gaussian pulses}\\
\frac{1}{2\pi}\approx0.16, & \text{for rectangular pulses}\\
\frac{2}{\pi}\approx0.64, & \text{for exponential pulses}
\end{cases}
\label{eq:Geometric_factor}
\end{align}

Therefore, by using these geometric factors, the QFI bounds can be written in terms of the pulse duration as

\begin{align}
    \mathcal{G}^{2}\mathcal{Q}_{\mathcal{G}}^{\pm}=&16\pi c_{\mathrm{P}}T\mathcal{G}^{2}\left[\frac{1}{2\mathcal{K}\mathcal{G}}\left(1\mp\frac{\Delta_{\mathrm{AC}}}{\Omega_{\mathrm{R}}}\right)-\frac{\pi c_{\mathrm{P}}T}{\Omega_{\mathrm{R}}^{2}}\right]\label{eq:polariton_bounds_QFI}
\end{align}

\section{CFI of TCSPC measurements\label{CFI_app}}

At asymptotic time, the probability of detecting a photon in the temporal mode $a\left(\tau\right)$ is given by

\begin{align}
p\left(\tau;\theta\right)= & \text{\ensuremath{\left|\psi_{\mathrm{out}}\left(\tau\right)\right|}}^{2}\label{eq:PD_probability}
\end{align}

Therefore, the CFI of estimating $\theta$ from a TCSPC measurement is given by

\begin{align}
\mathcal{C}_{\theta}^{\mathrm{T}}= & \int_{-\infty}^{+\infty} d\tau\frac{\left[\partial_{\theta}\left|\psi_{\mathrm{out}}\left(\tau\right)\right|^{2}\right]^{2}}{\left|\psi_{\mathrm{out}}\left(\tau\right)\right|^{2}}
\end{align}

the temporal output pulse shape is given by the  transform of the output pulse shape in the frequency domain $\psi_{\mathrm{out}}\left(\tau\right)=\mathcal{F}\left[\tilde{\psi}_{\mathrm{out}}\left(\omega\right)\right]=\int_{-\infty}^{-\infty}d\omega\tilde{\psi}_{\mathrm{out}}\left(\omega\right)\exp\left(-i\omega\tau\right)/\sqrt{2\pi}$. Since the output pulse in the frequency domain only is a unitary on the input pulse, we have

\begin{align}
\tilde{\psi}_{\mathrm{out}}\left(\omega\right)= & e^{-i\varphi\left(\omega\right)}\tilde{\psi}_{\mathrm{in}}\left(\omega\right)
\end{align}

The $\theta$-derivative of the probability in Eq.~\eqref{eq:PD_probability} is then given by

\begin{align}
p\left(\tau;\theta\right)= & \text{\ensuremath{\left|\psi_{\mathrm{out}}\left(\tau\right)\right|}}^{2}=\psi_{\mathrm{out}}^{*}\left(\tau\right)\psi_{\mathrm{out}}\left(\tau\right)\nonumber \\
\frac{\partial p\left(\tau;\theta\right)}{\partial\theta}= & \left\{ \partial_{\theta}\psi_{\mathrm{out}}^{*}\left(\tau\right)\right\} \psi_{\mathrm{out}}\left(\tau\right)+\psi_{\mathrm{out}}^{*}\left(\tau\right)\left\{ \partial_{\theta}\psi_{\mathrm{out}}\left(\tau\right)\right\} 
\end{align}

The $\theta$-derivative of the temporal pulse shape is therefore written in the frequency domain as

\begin{align}
\partial_{\theta}\psi_{\mathrm{out}}\left(\tau\right)= & \partial_{\theta}\mathcal{F}\left[\tilde{\psi}_{\mathrm{out}}\left(\omega\right)\right]=\mathcal{F}\left[\partial_{\theta}\tilde{\psi}_{\mathrm{out}}\left(\omega\right)\right]\nonumber \\
= & -i\mathcal{F}\left[\mathcal{Z}_{\theta}\left(\omega\right)\tilde{\psi}_{\mathrm{out}}\left(\omega\right)\right]=-i\chi\left(\tau\right)\label{eq:chi_function}
\end{align}

Since we are dealing only with functions of real variables and complex conjugation commutes with the differentiation operation, the $\theta$-derivative of the complex conjugate of the temporal pulse shape is given by

\begin{align}
\partial_{\theta}\psi_{\mathrm{out}}^{*}\left(\tau\right)= & \left[\partial_{\theta}\psi_{\mathrm{out}}\left(\tau\right)\right]^{*}=+i\chi^{*}\left(\tau\right)
\end{align}

Therefore we have 
\begin{align}
\frac{\partial}{\partial\theta}p\left(\tau;\theta\right)=\partial_{\theta}\text{\ensuremath{\left|\psi_{\mathrm{out}}\left(\tau\right)\right|}}^{2}= & +i\chi^{*}\left(\tau\right)\psi_{\mathrm{out}}\left(\tau\right)-i\chi\left(\tau\right)\psi_{\mathrm{out}}^{*}\left(\tau\right)\nonumber \\
= & 2\mathrm{Im}\left[\chi\left(\tau\right)\psi_{\mathrm{out}}^{*}\left(\tau\right)\right]
\end{align}

Therefore, the CFI of estimation of $\theta$ from a TCSPC measurement at asymptotic time is given by

\begin{align}
\mathcal{C}_{\theta}^{\mathrm{T}}= & 4\int_{-\infty}^{+\infty}d\tau\frac{\left\{ \mathrm{Im}\left[\chi\left(\tau\right)\psi_{\mathrm{out}}^{*}\left(\tau\right)\right]\right\} ^{2}}{\left|\psi_{\mathrm{out}}\left(\tau\right)\right|^{2}}\label{eq:TCSPC_CFI}
\end{align}

\subsection{Output Pulse in the time domain}

The expression in Eq.~\eqref{eq:freq_output_shape} tells us what is the frequency domain output pulse shape in terms of the cavity Green's function in the frequency domain. In terms of the shifted frequency coordinate $\Omega=\omega+\Delta$, this is written as

\begin{align}
\tilde{G}_{\mathrm{C}}\left(\Omega\right)= & -i\sqrt{\frac{\mathcal{K}}{2\pi}}\frac{\Omega}{g^{2}-\Omega^{2}-i\Omega\alpha}
\end{align}

The poles of this Green's function, which are found  at the zeros of the denominator, are calculated easily in terms of the shifted frequency coordinate $\Omega$. The roots of the polynomial in the denominator $D \left( \Omega \right)$ may be found as the  solutions to the equation $-D\left( \Omega \right)=0$, and are written as

\begin{align}
\Omega_{\pm}= & \frac{-i}{2}\left(\alpha\mp\Phi\right)
\end{align}

allowing us to write

\begin{align}
D\left(\Omega\right)=-\left(\Omega-\Omega_{+}\right)\left(\Omega-\Omega_{-}\right)    
\end{align}

and the cavity Green's function may be written in terms of these poles as

\begin{align}
\tilde{G}_{\mathrm{C}}\left(\Omega\right)= & i\sqrt{\frac{\mathcal{K}}{2\pi}}\frac{\Omega}{\left(\Omega-\Omega_{+}\right)\left(\Omega-\Omega_{-}\right)}
\end{align}

Using the partial fraction decomposition $\frac{\Omega}{\left(\Omega-\Omega_{+}\right)\left(\Omega-\Omega_{-}\right)}=\frac{A}{\left(\Omega-\Omega_{+}\right)}+\frac{B}{\left(\Omega-\Omega_{-}\right)}$ for $A=\frac{\Omega_{+}}{\Omega_{+}-\Omega_{-}}$ and $B=-\frac{\Omega_{-}}{\Omega_{+}-\Omega_{-}}$, the above maybe written as

\begin{align}
\tilde{G}_{\mathrm{C}}\left(\Omega\right)= & \frac{1}{\Phi}\sqrt{\frac{\mathcal{K}}{2\pi}}\left\{ \frac{\Omega_{+}}{\Omega-\Omega_{+}}-\frac{\Omega_{-}}{\Omega-\Omega_{-}}\right\} 
\end{align}

where we have used the difference $\Omega_{+}-\Omega_{-}=i\Phi$. Moving back to the original frequency coordinate $\omega$, we get

\begin{align}
\tilde{G}_{\mathrm{C}}\left(\omega\right)= & \frac{1}{\Phi}\sqrt{\frac{\mathcal{K}}{2\pi}}\left\{ \frac{\omega_{+}+\Delta}{\omega-\omega_{+}}-\frac{\omega_{-}+\Delta}{\omega-\omega_{-}}\right\} 
\end{align}

Using the above expression, the pulse Green's function or the response of the atom-cavity system maybe written as

\begin{align}
\tilde{G}_{\mathrm{P}}\left(\omega\right)= & 1-\frac{\mathcal{K}}{\Phi}\left\{ \frac{\omega_{+}+\Delta}{\omega-\omega_{+}}-\frac{\omega_{-}+\Delta}{\omega-\omega_{-}}\right\} 
\end{align}

Therefore the output pulse shape in the frequency domain is written as

\begin{align}
\tilde{\psi}_{\mathrm{out}}\left(\omega\right)= & \tilde{G}_{\mathrm{P}}\left(\omega\right)\tilde{\psi}_{\mathrm{in}}\left(\omega\right)\label{eq:TD_output_pulse}\\
= & \tilde{\psi}_{\mathrm{in}}\left(\omega\right)-\frac{\mathcal{K}}{\Phi}\left\{ \frac{\omega_{+}+\Delta}{\omega-\omega_{+}}-\frac{\omega_{-}+\Delta}{\omega-\omega_{-}}\right\} \tilde{\psi}_{\mathrm{in}}\left(\omega\right)\nonumber 
\end{align}

Performing an inverse Fourier transform allows us to obtain the output pulse in the time domain

\begin{align}
\psi_{\mathrm{out}}\left(t\right)= & \psi_{\mathrm{in}}\left(t\right)-\mathcal{F}^{-1}\left[\frac{\mathcal{K}}{\Phi}\left\{ \frac{\omega_{+}+\Delta}{\omega-\omega_{+}}-\frac{\omega_{-}+\Delta}{\omega-\omega_{-}}\right\} \tilde{\psi}_{\mathrm{in}}\left(\omega\right)\right]\label{eq:time_output_shape}
\end{align}
The above equation can be used to obtain the output temporal pulse shapes for any arbitrary input pulse shape.

\subsection{Output temporal pulse shape for an exponential input pulse}

Here we use Eq.~\eqref{eq:time_output_shape} to analytically calculate the expressions for the output temporal pulse shape obtained from an exponential input pulse. For a rising exponential pulse in the time domain, the frequency domain input pulse is given by
\begin{align}
\tilde{\psi}_{\mathrm{Exp}}\left(\omega\right)= & \sqrt{\frac{2T}{\pi}}\cdot\frac{1}{1+2i\omega T}
\end{align}

An arbitrary shear can be applied to this input pulse, which is equivalent to an additional phase in the time domain under the Fourier transform. Thus, applying this shear is equivalent to changing the carrier frequency of the input pulse. Under this shear, the input pulse can be written as follows

\begin{align}
\tilde{\psi}_{\mathrm{Exp}}\left(\omega+\omega_{0}\right)= & \sqrt{\frac{1}{2\pi T}}\frac{-i}{\left(\omega-\omega_{\mathrm{S}}\right)}
\end{align}

where we have

\begin{align}
\omega_{\mathrm{S}}= & -\omega_{0}+\frac{i}{2T}
\end{align}

We apply this input to the expression in Eq.~\eqref{eq:time_output_shape} and evaluate the Fourier transforms by solving the integrals using semi-circular contours closing in the upper and lower half planes for $\tau<0$ and $\tau>0$ respectively. This allows us to obtain

\begin{align}
\psi_{\mathrm{out}}\left(\tau<0\right)= & \frac{1}{\sqrt{T}}e^{i\omega_{0}\tau}e^{\frac{\tau}{2T}}\left(1-\frac{\mathcal{K}}{\Phi}\left[\mathcal{T}_{+}-\mathcal{T}_{-}\right]\right)\\
\psi_{\mathrm{out}}\left(\tau>0\right)= & -\frac{\mathcal{K}}{\Phi\sqrt{T}}\left( \mathcal{T}_{+}e^{+\frac{\Phi}{2}\tau}-\mathcal{T}_{-}e^{-\frac{\Phi}{2}\tau}\right) e^{-\frac{\mathcal{K}}{4}\tau}e^{+\frac{i}{2}\Delta\tau}
\end{align}

where we have defined the temporal amplitudes

\begin{align}
\mathcal{T}_{\pm} & =\frac{\omega_{\pm}+\Delta}{\omega_{\mathrm{S}}-\omega_{\pm}}
\end{align}

\begin{figure}[t]
    \centering
     
    (a)\includegraphics[scale=0.49]{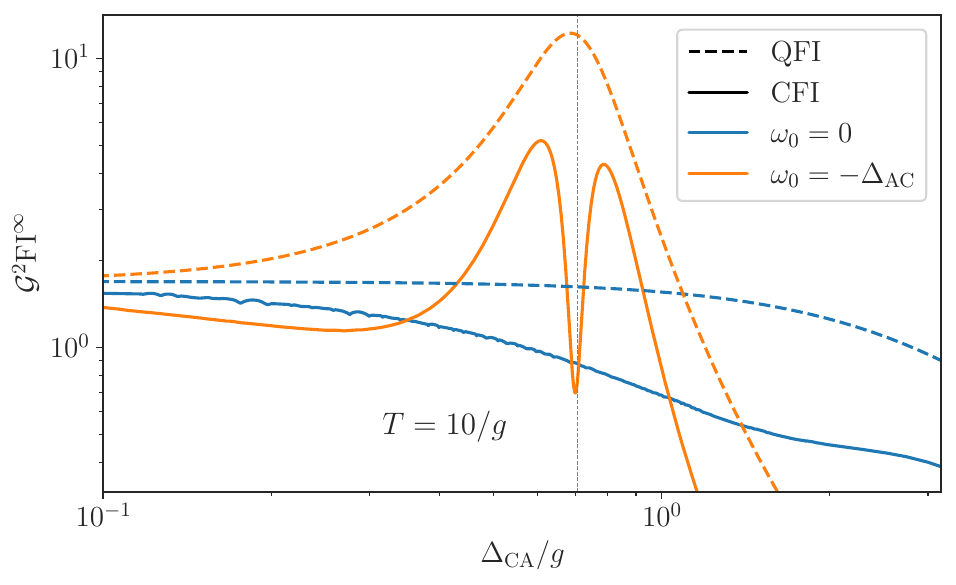} 
    (b)\includegraphics[scale=0.49]{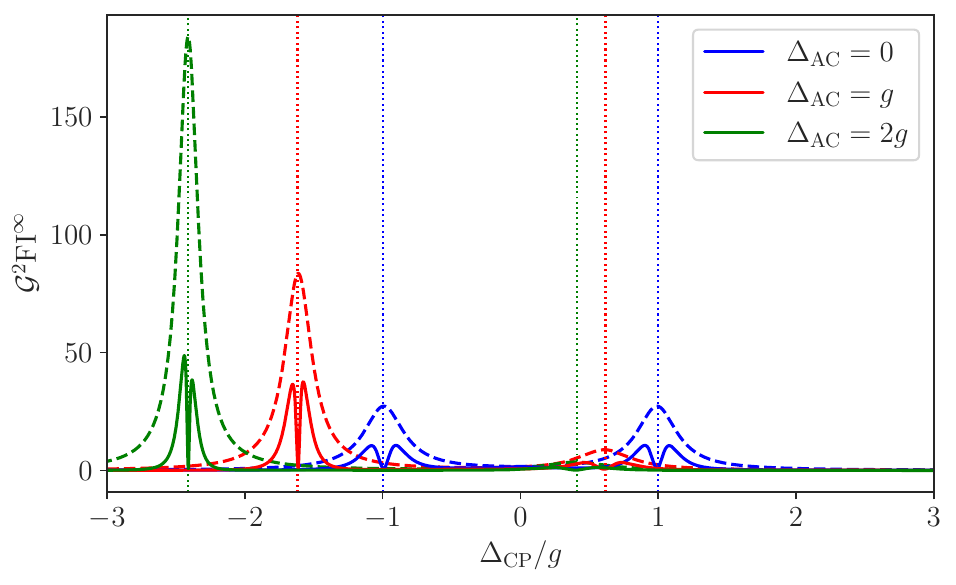}
    
    \caption{ The QFI (dashed lines) and CFI of TCSPC (solid lines) obtained for an exponentially rising single-photon pulse shape are shown. (a) The dimensionless FI values are plotted against the values of the atom-cavity detuning, scaled by the JC interaction strength. The grey vertical line is at the $\Delta_{\mathrm{AC}}=g/\sqrt{2}$. (b) The FI values are plotted against the values of the pulse-cavity detuning for different values of the atom cavity detuning. For these plots the value of $\mathcal{K}/g$ was set to 0.5. The value of the JC coupling strength was taken to be $g=3.3\times 10^{-4} \mathrm{cm}^{-1}\approx 10 \mathrm{MHz}$ }
    \label{fig:CFI_TCSPC}
\end{figure}

The denominators of these temporal mode amplitudes are complex in general and it can be shown that their real components can go to zero at certain values of the atom-cavity detuning and the input shear. The values correspond to resonances between the input shear and the polariton frequencies of the atom-cavity system. When the input shear is set such that $\omega_0=-\Delta$, this resonance condition is met at $\Delta=\frac{g}{\sqrt{2}}$. The CFI of TCPSC around this resonance point is shown in Fig.~\ref{fig:CFI_TCSPC}(a). It can be seen that for the long pulses considered, there is a significant enhancement obtained in the CFI when by detuning the atom-cavity system and tuning the carrier frequency of the input pulse by varying $\omega_0$. In Fig.~\ref{fig:CFI_TCSPC}(b) we show the DAE effect in the CFI more generally by plotting the FI values against the pulse-cavity detuning for different values of the detuning. We observe that the CFI peaks for $\Delta_{\mathrm{AC}}=2g$ are not only higher than the resonant CFI values, but also higher than the QFI values at resonance. Thus the DAE offers an enhancement which can be extracted by TCSPC and the enhancement allows us to exceed the ultimate precision limits at resonance.

\section{CES with pulsed single-photon probes in the lossy regime\label{SP_loss}}

The Hamiltonian consists of four components - atom, cavity, pulse and the environment modes.

\begin{align}
H & =H_{\mathrm{A}}+H_{\mathrm{C}}+H_{\mathrm{P}}+H_{\mathrm{E}}+H_{\mathrm{Int}}
\end{align}

The free part of the environment Hamiltonian is written as

\begin{align}
H_{\mathrm{E}} & =\sum_{k\in\{1,2\}}\int_{0}^{\infty}\omega b_{k}^{\dag}\left(\omega\right)b_{k}\left(\omega\right)d\omega
\end{align}

where the two sets of the environment modes correspond to modes which couple to the atom and the cavity respectively. The interaction Hamiltonian has the following parts

\begin{align}
H_{\mathrm{Int}}= & H_{\mathrm{AC}}+H_{\mathrm{CP}}+H_{\mathrm{AE}}+H_{\mathrm{CE}}
\end{align}

The first two interactions are the same as the lossless case. The last two interaction terms represent the coupling of the atom and the cavity to the environment. Here we have also assumed that the incident travelling probe field interacts with the atom only through the selected cavity mode. Within this model, direct coupling between the incident probe and the atom is not considered. Coupling of the atom to other electromagnetic modes other than the cavity mode is represented by the atomic loss rate $\Gamma$. This separates the measured cavity input--output field from the unobserved electromagnetic modes responsible for atomic loss. Similar to how we proceed in the lossless case, we first move to the interaction picture wrt the free Hamiltonian $H_{0}=H_{\mathrm{A}}+H_{\mathrm{C}}+H_{\mathrm{P}}+H_{\mathrm{E}}$. In this picture, we make the slowly varying envelope approximation by assuming that the light in the waveguide and the environments are narrowband about some particular frequencies and introduce the following white noise operators for the pulse and the environment modes. 

\begin{align}
\tilde{a}\left(t\right)= & \frac{1}{\sqrt{2\pi}}\int_{-\infty}^{+\infty}a\left(\Omega+\omega_{\mathrm{P}}\right)e^{-i\Omega t}d\Omega\\
b_{1}\left(t\right)= & \frac{1}{\sqrt{2\pi}}\int_{-\infty}^{+\infty}b_{1}\left(\Omega+\omega_{\mathrm{A}}\right)e^{-i\Omega t}d\Omega\\
b_{2}\left(t\right)= & \frac{1}{\sqrt{2\pi}}\int_{-\infty}^{+\infty}b_{2}\left(\Omega+\omega_{\mathrm{C}}\right)e^{-i\Omega t}d\Omega
\end{align}

By introducing $a\left(t\right)= e^{-i\Delta_{\mathrm{CP}}t}\tilde{a}\left(t\right)$, the total Hamiltonian in the interaction picture is written in terms of these white noise operators as

\begin{align}
H_{\mathrm{I}}^{\mathrm{ACPE}}= & -i\hbar\left(ge^{-i\Delta_{\mathrm{AC}}t}\sigma^{\dag}c+\sqrt{\mathcal{K}}ca^{\dag}\left(t\right)+\sqrt{\Gamma}\sigma^{\dag}b_{1}\left(t\right)+\sqrt{\mathcal{K}_{\perp}}cb_{2}^{\dag}\left(t\right)+\mathrm{h.c}\right)
\label{eq:lossy_H_int}
\end{align}

We take the following ansatz to solve this

\begin{align}
\left|\Psi^{\mathrm{ACPE}}\left(t\right)\right\rangle = & \psi_{\mathrm{A}}\left(t\right)\left|e,0_{\mathrm{C}},0_{\mathrm{P}},0_{\mathrm{E}}\right\rangle +\psi_{\mathrm{C}}\left(t\right)\left|g,1_{\mathrm{C}},0_{\mathrm{P}},0_{\mathrm{E}}\right\rangle \\
+ & \left|g,0_{\mathrm{C}}\right\rangle \otimes\left\{ \left|\tilde{\psi}_{\mathrm{P}}\left(t\right)\right\rangle \otimes\left|0_{\mathrm{E}}\right\rangle +\left|0_{\mathrm{P}}\right\rangle \otimes\left[\left|\tilde{\psi}_{\mathrm{E}}^{1}\left(t\right)\right\rangle \otimes\left|0_{\mathrm{E}}^{2}\right\rangle +\left|0_{\mathrm{E}}^{1}\right\rangle \otimes\left|\tilde{\psi}_{\mathrm{E}}^{2}\left(t\right)\right\rangle \right]\right\} 
\end{align}

By substituting the above ansatz into the Schrodinger equation $i\hbar\partial_{t}\left|\Psi^{\mathrm{ACPE}}\left(t\right)\right\rangle =H_{\mathrm{I}}^{\mathrm{ACPE}}\left(t\right)\left|\Psi^{\mathrm{ACPE}}\left(t\right)\right\rangle $ we obtain the following dynamical equations

\begin{align}
\frac{\partial}{\partial t}\psi_{\mathrm{A}}\left(t\right)= & -ge^{-i\Delta_{\mathrm{AC}}t}\psi_{\mathrm{C}}\left(t\right)-\frac{\Gamma}{2}\psi_{\mathrm{A}}\left(t\right)\label{eq:psi_A_CDE}\\
\frac{\partial}{\partial t}\psi_{\mathrm{C}}\left(t\right)= & +ge^{+i\Delta_{\mathrm{AC}}t}\psi_{\mathrm{A}}\left(t\right)+\sqrt{\mathcal{K}}\xi_{\mathrm{in}}\left(t\right)-\frac{\mathcal{K}_{\mathrm{tot}}}{2}\psi_{\mathrm{C}}\left(t\right)\label{eq:psi_C_CDE}\\
\frac{\partial}{\partial t}\left|\tilde{\psi}_{\mathrm{P}}\left(t\right)\right\rangle = & -\sqrt{\mathcal{K}}\psi_{\mathrm{C}}\left(t\right)a^{\dag}\left(t\right)\left|0_{\mathrm{P}}\right\rangle \\
\frac{\partial}{\partial t}\left|\tilde{\psi}_{\mathrm{E},1}\left(t\right)\right\rangle = & +\sqrt{\Gamma}\psi_{\mathrm{A}}\left(t\right)b_{1}^{\dag}\left(t\right)\left|0_{\mathrm{E},1}\right\rangle \\
\frac{\partial}{\partial t}\left|\tilde{\psi}_{\mathrm{E},2}\left(t\right)\right\rangle = & -\sqrt{\mathcal{K}_{\perp}}\psi_{\mathrm{C}}\left(t\right)b_{2}^{\dag}\left(t\right)\left|0_{\mathrm{E},2}\right\rangle 
\end{align}

These dynamical equations can be solved by following the method outlined in~\cite{Banacloche_2016} and yield the following solutions.

\begin{align}
\psi_{\mathrm{C}}\left(t\right)= & \frac{\sqrt{\mathcal{K}}}{2\Omega_{\mathrm{eff}}}\int_{-\infty}^{t}d\tau\sum_{s=\pm}s\left(\frac{\Gamma_{\mathrm{eff}}}{2}+s\Omega_{\mathrm{eff}}\right)e^{\lambda_{\mathrm{C},s}\left(t-\tau\right)}\xi_{\mathrm{in}}\left(\tau\right)\\
\psi_{\mathrm{A}}\left(t\right)= & -\frac{g\sqrt{\mathcal{K}}}{2\Omega_{\mathrm{eff}}}\int_{-\infty}^{t}d\tau\sum_{s=\pm}se^{\lambda_{\mathrm{A},s}\left(t-\tau\right)}e^{-i\Delta_{\mathrm{AC}}\tau}\xi_{\mathrm{in}}\left(\tau\right)\\
\left|\tilde{\psi}_{\mathrm{P}}\left(t\right)\right\rangle = & \int_{-\infty}^{+\infty}d\tau\left(\xi_{\mathrm{in}}\left(\tau\right)-\sqrt{\mathcal{K}}\Theta\left(t-\tau\right)\psi_{\mathrm{C}}\left(\tau\right)\right)a^{\dag}\left(\tau\right)\left|0_{\mathrm{P}}\right\rangle 
\end{align}

where the parameters are defined as

\begin{align*}
\Gamma_{\mathrm{eff}}= & \frac{\Gamma-\mathcal{K}_{\mathrm{tot}}}{2}-i\Delta_{\mathrm{AC}} & \lambda_{\mathrm{C},\pm}= & -\frac{\Gamma_{\mathrm{tot}}}{2}\pm\Omega_{\mathrm{eff}}\\
\mathcal{K}_{\mathrm{tot}}= & \mathcal{K}+\mathcal{K}_{\perp} & \lambda_{\mathrm{A},\pm}= & -\frac{\Gamma_{\mathrm{tot}}^{*}}{2}\pm\Omega_{\mathrm{eff}}\\
\Gamma_{\mathrm{tot}}= & \frac{\mathcal{K}_{\mathrm{tot}}+\Gamma}{2}-i\Delta_{\mathrm{AC}} & \Omega_{\mathrm{eff}}= & \frac{\sqrt{\Gamma_{\mathrm{eff}}^{2}-4g^{2}}}{2}
\end{align*}

The states of the environment fields are written as
\begin{align}
\left|\tilde{\psi}_{\mathrm{E},1}\left(t\right)\right\rangle = & \left|0_{\mathrm{E},1}\right\rangle +\sqrt{\Gamma}\int_{-\infty}^{t}d\tau\psi_{\mathrm{A}}\left(\tau\right)b_{1}^{\dag}\left(\tau\right)\left|0_{\mathrm{E},1}\right\rangle \\
\left|\tilde{\psi}_{\mathrm{E},2}\left(t\right)\right\rangle = & \left|0_{\mathrm{E},2}\right\rangle -\sqrt{\mathcal{K}_{\perp}}\int_{-\infty}^{t}d\tau\psi_{\mathrm{C}}\left(\tau\right)b_{2}^{\dag}\left(\tau\right)\left|0_{\mathrm{E},2}\right\rangle 
\end{align}

The total state of the density matrix is written as
\begin{align}
\rho^{\mathrm{ACPE}}\left(t\right)= & \left|\Psi^{\mathrm{ACPE}}\left(t\right)\right\rangle \left\langle \Psi^{\mathrm{ACPE}}\left(t\right)\right|
\end{align}

The state of the pulse can be obtained by tracing out the other subsystems as 

\begin{align}
\rho^{\mathrm{P}}\left(t\right)= & \mathrm{Tr}_{\mathrm{ACE}}\left(\rho^{\mathrm{ACPE}}\left(t\right)\right)
\end{align}

\begin{figure}[t]
    \centering
    \includegraphics[width=\linewidth]{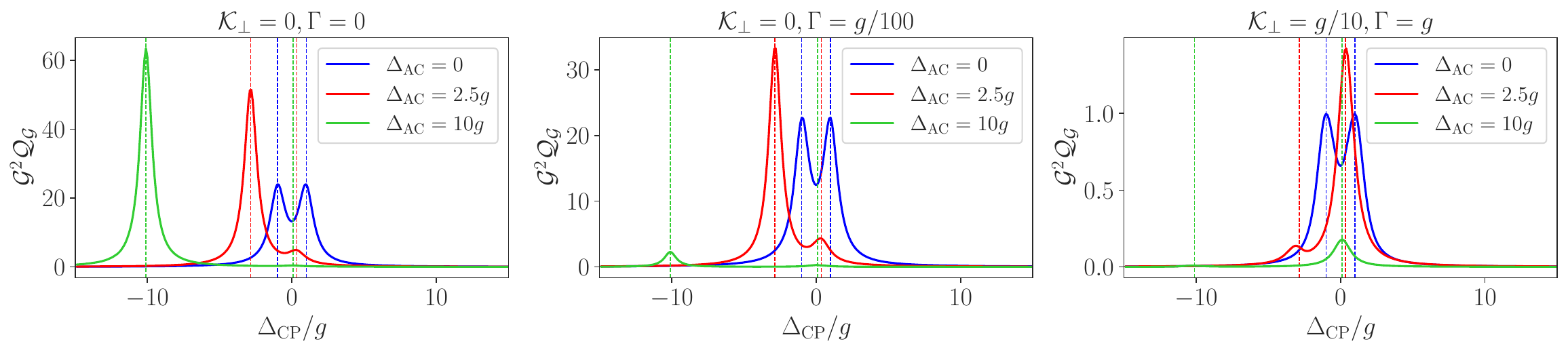}

    \caption{Dimensionless QFI for the single-photon exponential pulse plotted against the pulse-cavity detuning $\Delta_{\mathrm{CP}}$ for different values of $\Delta_{\mathrm{AC}}/g$ and loss parameters $\mathcal{K}_{\perp}$and $\Gamma$. The colors correspond to different values of the atom-cavity detuning $\Delta_{\mathrm{AC}}$. The vertical black lines correspond to the two polariton frequencies. The value of the JC coupling strength was taken to be $g=3.3\times 10^{-4} \mathrm{cm}^{-1}\approx 10 \mathrm{MHz}$. The cavity-pulse coupling and the pulse duration were taken to be $\mathcal{K}_{\perp}=g/2$ and $T=1/g$ respectively.\label{fig:lossy_QFI}}
    
\end{figure}

It is easy to see that this state may be written as

\begin{align}
\rho^{\mathrm{P}}\left(t\right)= & \left(1-p_{\mathrm{P}}\left(t\right)\right)\left|0_{\mathrm{P}}\right\rangle \left\langle 0_{\mathrm{P}}\right|+\left|\tilde{\psi}_{\mathrm{P}}\left(t\right)\right\rangle \left\langle \tilde{\psi}_{\mathrm{P}}\left(t\right)\right|
\end{align}

where $p_{\mathrm{P}}\left(t\right)=\left\langle \tilde{\psi}_{\mathrm{P}}\left(t\right)\left|\tilde{\psi}_{\mathrm{P}}\left(t\right)\right\rangle \right.$ is the probability of the excitation being in the pulse modes. For an input pulse shape $\xi_{\mathrm{in}}\left(\tau\right)$, the input and the output pulse at asymptotic time is then given by

\begin{align}
\left|\tilde{\psi}_{\mathrm{P,out}}\right\rangle = & \int_{-\infty}^{+\infty}d\tau\xi_{\mathrm{in}}\left(\tau\right)a^{\dag}\left(\tau\right)\left|0_{\mathrm{P}}\right\rangle -\sqrt{\mathcal{K}}\int_{-\infty}^{+\infty}d\tau\psi_{\mathrm{C}}\left(\tau\right)a^{\dag}\left(\tau\right)\left|0_{\mathrm{P}}\right\rangle 
\end{align}

The probability of the photon being in the output pulse is $p_{\mathrm{P}}^{\mathrm{out}}=\left\langle \tilde{\psi}_{\mathrm{P,out}}\left|\tilde{\psi}_{\mathrm{P,out}}\right\rangle \right.$. We can re-write the output pulse state, $\rho_{\mathrm{out}}^{\mathrm{P}}=\left(1-p_{\mathrm{P}}^{\mathrm{out}}\right)\left|0_{\mathrm{P}}\right\rangle \left\langle 0_{\mathrm{P}}\right|+\left|\tilde{\psi}_{\mathrm{P,out}}\right\rangle \left\langle \tilde{\psi}_{\mathrm{P,out}}\right|$, in terms of the normalised state

\begin{align}
\left|\psi_{\mathrm{P}}^{\mathrm{out}}\right\rangle = & \frac{\left|\tilde{\psi}_{\mathrm{P,out}}\right\rangle }{\sqrt{\left\langle \tilde{\psi}_{\mathrm{P,out}}\left|\tilde{\psi}_{\mathrm{P,out}}\right\rangle \right.}}=\frac{\left|\tilde{\psi}_{\mathrm{P,out}}\right\rangle }{\sqrt{p_{\mathrm{P}}^{\mathrm{out}}}}
\end{align}

Thus we have  $\left|\tilde{\psi}_{\mathrm{P,out}}\right\rangle \left\langle \tilde{\psi}_{\mathrm{P,out}}\right|=  p_{\mathrm{P}}^{\mathrm{out}}\left|\psi_{\mathrm{P}}^{\mathrm{out}}\right\rangle \left\langle \psi_{\mathrm{P}}^{\mathrm{out}}\right|$. In terms of this normalised state, the density matrix is given by

\begin{align}
\rho_{\mathrm{out}}^{\mathrm{P}}= & p_{0}\left|0_{\mathrm{P}}\right\rangle \left\langle 0_{\mathrm{P}}\right|+p_{1}\left|\psi_{\mathrm{P}}^{\mathrm{out}}\right\rangle \left\langle \psi_{\mathrm{P}}^{\mathrm{out}}\right|
\end{align}

where we have redefined  $p_1=p_{\mathrm{P}}^{\mathrm{out}}$ and $p_0=1-p_{\mathrm{P}}^{\mathrm{out}}$. It is easy to see from the expression for the output pulse state that the output pulse shape maybe written as a convolution of a Green's function with the input pulse shape in the time domain $\xi_{\mathrm{out}}\left(\tau\right)=  \xi_{\mathrm{in}}\left(\tau\right)-\mathcal{K}\int_{-\infty}^{\tau}d\tau_{1}G_{\mathrm{C}}\left(\tau-\tau_{1}\right)\xi_{\mathrm{in}}\left(\tau_{1}\right)$, where

\begin{align}
G_{\mathrm{C}}\left(\tau\right)= & \frac{1}{2\Omega_{\mathrm{eff}}}\sum_{s=\pm}s\left(\frac{\Gamma_{\mathrm{eff}}}{2}+s\Omega_{\mathrm{eff}}\right)e^{\lambda_{\mathrm{C},\pm}\tau}
\end{align}

Under a Fourier transform $\tilde{f}\left(\omega\right)=\int_{-\infty}^{\infty}dtf\left(t\right)e^{i\omega t}/\sqrt{2\pi}$, we can move to the frequency domain, where the convolution becomes a product.

\begin{align}
\tilde{\xi}_{\mathrm{out}}\left(\omega\right)= & \tilde{\xi}_{\mathrm{in}}\left(\omega\right)-\sqrt{2\pi}\mathcal{K}\tilde{G}_{\mathrm{C}}\left(\omega\right)\tilde{\xi}_{\mathrm{in}}\left(\omega\right)=S\left(\omega\right)\tilde{\xi}_{\mathrm{in}}\left(\omega\right)
\end{align}

where $\tilde{\xi}$ represents the Fourier transformed pulse shapes and $\tilde{G}_{\mathrm{C}}$ is the Fourier transform of the time domain Green's function. We have also defined the scattering coefficient $S\left(\omega\right)=1-\sqrt{2\pi}\mathcal{K}\tilde{G}_{\mathrm{C}}\left(\omega\right)$. The derivative of the output spectral amplitude is

\begin{align}
\partial_{\mathcal{G}}\tilde{\xi}_{\mathrm{out}}\left(\omega\right)= & \partial_{\mathcal{G}}S\left(\omega\right)\tilde{\xi}_{\mathrm{in}}\left(\omega\right)
= S_{\mathcal{G}}\left(\omega\right)\tilde{\xi}_{\mathrm{in}}\left(\omega\right)
\end{align}

The scattered coefficient and its $\mathcal{G}$-derivative are expressed in terms of the system parameters as
\begin{align}
S\left(\omega\right)= & 1+\frac{\mathcal{K}}{2\Omega_{\mathrm{eff}}}\left\{ \frac{\frac{\Gamma_{\mathrm{eff}}}{2}+\Omega_{\mathrm{eff}}}{\Omega_{\mathrm{eff}}-\frac{\Gamma_{\mathrm{tot}}}{2}+i\omega}+\frac{\frac{\Gamma_{\mathrm{eff}}}{2}-\Omega_{\mathrm{eff}}}{\left(\Omega_{\mathrm{eff}}+\frac{\Gamma_{\mathrm{tot}}}{2}-i\omega\right)}\right\}\\S_{\mathcal{G}}\left(\omega\right)= & \frac{\mathcal{K}}{4\Omega_{\mathrm{eff}}^{3}}\left\{ \frac{\frac{\Gamma_{\mathrm{eff}}}{2}\left(\lambda_{\mathrm{C}}^{+}+i\omega\right)+\varUpsilon_{+}}{\left(\lambda_{\mathrm{C}}^{+}+i\omega\right)^{2}}-\frac{\frac{\Gamma_{\mathrm{eff}}}{2}\left(\lambda_{\mathrm{C}}^{-}+i\omega\right)+\varUpsilon_{-}}{\left(\lambda_{\mathrm{C}}^{-}+i\omega\right)^{2}}\right\} 
\end{align}

where we have defined the parameters  $\varUpsilon_{\pm}=\Omega_{\mathrm{eff}}\left(\Omega_{\mathrm{eff}}\pm\frac{\Gamma_{\mathrm{eff}}}{2}\right)$. The asymptotic QFI of estimating $\mathcal{G}$ is expressed as

\begin{align}
\mathcal{Q}_{\mathcal{G}}\left(\rho_{\mathrm{out}}^{\mathrm{P}}\right)= & \frac{\left(\partial_{\mathcal{G}}p_1\right)^{2}}{p_0p_1}+4p_1\left\{ \left\langle \partial_{\mathcal{G}}\psi_{\mathrm{P}}^{\mathrm{out}}\left|\partial_{\mathcal{G}}\psi_{\mathrm{P}}^{\mathrm{out}}\right\rangle \right.-\left|\left\langle \partial_{\mathcal{G}}\psi_{\mathrm{P}}^{\mathrm{out}}\left|\psi_{\mathrm{P}}^{\mathrm{out}}\right\rangle \right.\right|^{2}\right\} 
\end{align}

While the above expression is written in terms of the normalised output state, the QFI can also be expressed in terms of the unnormalised states as

\begin{align}
\mathcal{Q}_{\mathcal{G}}\left(\rho_{\mathrm{out}}^{\mathrm{P}}\right)= & \frac{\left(\partial_{\mathcal{G}}p_{\mathrm{P}}^{\mathrm{out}}\right)^{2}}{\left(1-p_{\mathrm{P}}^{\mathrm{out}}\right)p_{\mathrm{P}}^{\mathrm{out}}}+4\left\{ \left\langle \partial_{\mathcal{G}}\tilde{\psi}_{\mathrm{P}}^{\mathrm{out}}\left|\partial_{\mathcal{G}}\tilde{\psi}_{\mathrm{P}}^{\mathrm{out}}\right\rangle \right.-\frac{\left|\left\langle \partial_{\mathcal{G}}\tilde{\psi}_{\mathrm{P}}^{\mathrm{out}}\left|\tilde{\psi}_{\mathrm{P}}^{\mathrm{out}}\right\rangle \right.\right|^{2}}{p_{\mathrm{P}}^{\mathrm{out}}}\right\} 
\end{align}

The unnormalised output state can be written in the frequency domain as $\left|\tilde{\psi}_{\mathrm{P}}^{\mathrm{out}}\right\rangle =\int d\omega S\left(\omega\right)\tilde{\xi}_{\mathrm{in}}\left(\omega\right)a^{\dag}\left(\omega+\omega_{\mathrm{C}}\right)\left|0_{\mathrm{P}}\right\rangle $. Therefore the components of the above expression can be computed in terms of the scattering coefficient and its derivative. Their expressions are as follows

\begin{align}
p_1= &  \int_{-\infty}^{+\infty}d\omega\left|S\left(\omega\right)\tilde{\xi}_{\mathrm{in}}\left(\omega\right)\right|^{2}\\
\partial_{\mathcal{G}}p_1= & 2\int_{-\infty}^{+\infty}d\omega\left|\tilde{\xi}_{\mathrm{in}}\left(\omega\right)\right|^{2}\Re\left[S^{*}\left(\omega\right)S_{\mathcal{G}}\left(\omega\right)\right]
\end{align}

Similarly, we have

\begin{align}
\left\langle \partial_{\mathcal{G}}\tilde{\psi}_{\mathrm{P}}^{\mathrm{out}}\left|\tilde{\psi}_{\mathrm{P}}^{\mathrm{out}}\right\rangle \right.= & \int_{-\infty}^{+\infty}d\omega S_{\mathcal{G}}^{*}\left(\omega\right)S\left(\omega\right)\left|\tilde{\xi}_{\mathrm{in}}\left(\omega\right)\right|^{2}\\
\left\langle \partial_{\mathcal{G}}\tilde{\psi}_{\mathrm{P}}^{\mathrm{out}}\left|\partial_{\mathcal{G}}\tilde{\psi}_{\mathrm{P}}^{\mathrm{out}}\right\rangle \right.= & \int_{-\infty}^{+\infty}d\omega\left|S_{\mathcal{G}}\left(\omega\right)\tilde{\xi}_{\mathrm{in}}\left(\omega\right)\right|^{2}
\end{align}

The QFI values computed using these expressions have been plotted in Fig.~\ref{fig:lossy_QFI} and show qualitatively similar behaviour to the photo-counting FIs plotted for the CW monochromatic probes. Thus, DAE at the level of pulsed single-photons persists in the presence of losses.

\end{document}